\documentclass[12pt,preprint]{aastex}
\slugcomment{ }

\shorttitle{Evolution of Habitable Zone and Search for Life}
\shortauthors{Danchi \& Lopez}

\begin{document}

\title{Effect of Metallicity on the Evolution of the Habitable Zone from the Pre-Main Sequence to the Asymptotic Giant Branch and the Search for Life}

\author{William C. Danchi}
\affil{Exoplanets and Stellar Astrophysics Laboratory, NASA Goddard Space Flight Center, \\ Code 667, Greenbelt, MD 20771 USA}
\email{william.c.danchi@nasa.gov}
\and 
\author{Bruno Lopez}
\affil{Observatoire de la C\^{o}te d'Azur, Laboratoire  Lagrange UMR 7293, BP 4229, \\ F-06034 Nice Cedex 4, France}
\email{bruno.lopez@oca.eu}

\begin{abstract}

During the course of stellar evolution, the location and width of the habitable zone changes as the  luminosity and radius of the star evolves.  The duration of habitability for a planet located at a given distance from a star is greatly affected by the characteristics of the host star.  A quantification of these effects can be used observationally in the search for life around nearby stars.  The longer the duration of habitability, the more likely it is that life has evolved.   The preparation of observational techniques aimed at detecting life would benefit from the scientific requirements deduced from the evolution of the habitable zone.   
We present a study of the evolution of the habitable zone around stars of 1.0, 1.5, and 2.0 M$_{\odot}$  for metallicities ranging from  Z=0.0001 to Z=0.070.  We also consider the evolution of the habitable zone  from the pre-main sequence until the asymptotic giant branch is reached.   
We find that metallicity strongly affects the duration of the habitable zone for a planet as well as the distance from the host star where the duration is maximized.  For a 1.0 M$_{\odot}$  star with near Solar metallicity, Z=0.017,  the duration of the habitable zone is $>$10 Gyr at distances  1.2 to 2.0 AU from the star, whereas the duration is $>$20 Gyr for high metallicity stars (Z=0.070) at distances of 0.7 to 1.8 AU, and  $\sim$4 Gyr at distances of 1.8 to 3.3 AU for low metallicity stars (Z=0.0001).   Corresponding results have been obtained for stars of 1.5 and 2.0 solar masses.

\end{abstract}


\section{Introduction}

As stars evolve from pre-main sequence (PMS)  to the main sequence and then to post-main sequence,
their luminosity evolves considerably.  For a 1 solar mass star,  at the beginning of the PMS the luminosity is as high as 10 solar luminosities.  A few million years later the luminosity decreases to a minimum of $\sim$ 0.4 solar luminosities, then it gradually rises to  $\sim$ 3 $L_{\odot}$ towards  the end of the main sequence, and it reaches a peak of more than a thousand  $L_{\odot}$ at the tip of the Asymptotic Giant Branch (AGB).   As a consequence of the evolution of the luminosity of a star, the location and the width of the habitable zone around the star also changes.  The habitable zone is defined to be an annulus or shell around the star in which liquid water can exist (Kasting, Whitmire, and Reynolds 1993).  For exoplanets located within the habitable zone the chance for life to develop depends on the length of time of its transit or its duration. 

In the past few years, the discovery of the planetary system around the nearby M dwarf, Gl581 (Udry et al. 2007, Mayor et al. 2009, Vogt et al. 2010), has generated increased interest and study of exoplanetary habitability because two of its exoplanets Gl581c and Gl581d are potentially habitable super-Earths.   Calculations of Selsis et al. (2007), and von Bloh (2007)  have shown that Gl581c is likely to be too close to its host star to be habitable and likely would suffer from a runaway greenhouse effect like Venus.  On the other hand, Gl581d, with a minimum mass of about 7 $M_{Earth}$ could be a habitable super-Earth if it had a minimum partial pressure of about 7 bar of $\mathrm{CO_2}$ with a total surface pressure of about 7.6 bar (Kaltenegger et al. 2011).  Similar conclusions concerning the habitability of Gl581d have been reached in studies by Wordsworth et al. (2011) and Hu and Ding (2011) with somewhat higher and lower partial pressures of $\mathrm{CO_2}$ than found in the work of Kaltenegger et al. (2011).   

A broader study of habitability of exoplanets is presented in the recent review paper by Lammer et al. (2009), who examined the important factors relevant to the evolution and habitability of Earth-like planets, including radiation and particle fluxes, geophysical environments and processes such as plate tectonics, the origin and evolution of life and its observability. 

At the present time, we do not know whether life evolved quickly  ---  within a few million or a hundred million years --- or much longer ---  up to a billion years, due to the difficulty of searching so far back into the Earth's fossil record.  Life on Earth could have been created and have evolved quickly, but could have been destroyed and then restarted during  the period of late-heavy bombardment.  Isotopic data have shown that life existed as early as $7 \times 10^8$ years after the Earth was formed (Mojzsis et al. 1997, Holland 1997).   Assuming that life evolved on the Earth on a relatively short time scale, Lineweaver and Davis (2002)  calculated that the probability of biogenesis around other planets could be $>$13\%  (95\% confidence level) for exoplanets in the habitable zones of their host stars, if they are older than 1 Gyr.  These authors use this statistical result to infer that life in the Universe may be common.  In a paper with results complementary to those of Lineweaver and Davis (2002), Davies (2003) hypothesizes that the rapid development of life on Earth could be due to a non-terrestrial origin, if life requires one or several improbable or ``hard'' steps, as discussed by  Carter (1983, 2008) and Hanson (1998).  Davies (2003) suggests that life could have developed on Mars and then transported to Earth, but is likely not of extrasolar origin.  

The hard-step scenario for the evolution of intelligent  life discussed in  Hanson (1998) and Carter (2008)  yields estimates for the mean time interval between the key steps.   The number of steps discussed in the literature varies between two (Carter 1983) and ten (Barrow \& Tipler 1986), and the mean interval between steps can be interpreted in the context of major evolutionary stages, such as biogenesis, bacterial, eukaryotes, combigenesis, metazoans, and civilization (Carter 2008).  The boundary between the oxygen-poor Archean eon and the oxygen-medium Proterozoic eon is at the third step, the eurokaryotes, which occurred around 2.3 Gyr ago.  The mean interval between steps is estimated to be about 0.8 Gyr (Carter 2008).  These two time intervals have observational consequences.  

By studying the evolution of the habitable zone around stars of different masses, metallicities, and evolutionary states  we have the possibility of empirically determining the time scale for life formation, at least in the sense of learning the time scale for life to affect the composition of the atmosphere of an exoplanet.   
 
Observationally it is well established that exoplanets with masses larger than Neptune mass (i.e., gas giants) are detected more frequently around higher-than-solar metallicity stars in the solar neighborhood from radial velocity surveys (Fischer and Valenti, 2005; Sousa et al. 2008).  However, this correlation does not appear to exist for lower mass exoplanets as shown recently by transit observations from Kepler (Buchhave et al. 2012).   In this case the distribution is approximately constant with metallicity for host stars with [Fe/H] = $-$0.5 to $+$0.5.  The correlation of gas giant exoplanets with high metallicity host stars supports the core accretion model of planet formation because their primordial disks are expected to have more rocky planetesimals from which the rocky cores of gas giants are formed (Ida \& Lin et al. 2004).   Recently, attention has been given to the effect of metallicity on the habitable zones around cool dwarfs  with masses  $<0.5~M_{\odot}$ with Kepler Objects of Interest (Muirhead et al. 2012) and the effect of abundance ratios of metals other than Fe in the evolution of the habitable zones around 1 $M_{\odot}$ stars (Young, Liebst, and Pagano 2012). 

Transit spectroscopy has been employed to characterize the physical conditions of exoplanets, including the discovery of a variety of molecular species around some close-in exoplanets (Deming et al. 2005, Richardson et al. 2007).  In the long run, imaging techniques such as coronagraphy at optical wavelengths and interferometry in the mid-infrared can be used in connection with spectroscopy to characterize the atmospheres of nearby exoplanets, not just transiting systems, which comprise only a few percent of observable systems.  

Research on the location and time scales for habitability around stars as they evolve from the main sequence has continued since the early work of Lopez et al. (2000), who first considered the search for planets and the location of the habitable zone around Red Giant Branch and Asymptotic Giant Branch stars in the context of the Darwin mission.  Later, Stern (2003) discussed the evolution of the habitable zone of 1-3 $M_{\odot}$ stars in the context of the possibility habitability of outer solar system bodies during the RGB and AGB phases, when the stars are at their luminosity peak.  

In a previous paper (Lopez, Schneider, \& Danchi 2005), hereafter Paper I, we studied the evolution of the habitable zone around stars of 1.0, 1.5, and 2.0 $M_{\odot}$ from the main sequence until the first ascent along the Red Giant Branch (RGB) for solar metallicity stars using the evolutionary tracks of Maeder and Meynet (1988).   In Paper I we showed that there is a second chance for life as a star evolves off the main sequence, because this evolution is very slow, particularly for stars near 1 $M_{\odot}$.  If life could develop and evolve over time scales from  about $5 \times 10^8$ to $10^9$ years, then there could be habitable planets around red giant stars.  For a 1 $M_{\odot}$ star, the duration of habitability was estimated to be $10^9$ years at 2 AU and around $10^8$ years at 9 AU.   This gives a second chance for life in the far distant future for planets such as Mars, which is currently too cold for liquid water to exist on its surface.  In this paper we review and show that after the first ascent along the RGB and the He flash, there is an additional period of time in which the luminosity is very stable, i.e., the core He burning phase, which can last a few times $10^8$ years or perhaps up to $10^9$ years.   During the core He burning phase there is also a potential for long duration habitability at distances of 7 to 22 AU.  

More recently, von Bloh et al. (2009) considered the evolution of the habitable zone and the duration of habitability, including not only the luminosity evolution of the host star but also the evolution of the atmospheric content of the planet itself, e.g. including the greenhouse effect and other processes.  The effects they included increased the habitable zone width by a small amount.  In the end, their results were quite similar to our own in terms of the duration of habitability as a star evolves off the main sequence.

In the present paper we show also that metallicity plays an important role in the evolution of the habitable zone. Stars having significantly higher-than-solar metallicity have a much longer duration of habitability at a given distance than do stars of lower than solar metallicity.  The difference in luminosity means that high metallicity stars have habitable zones that are closer  to the star than do low metallicity stars.  Thus strategies for the detection of life are strongly affected by the location of the habitable zone because all direct exoplanet detection and atmospheric characterization impose some specific requirements in terms of  spatial resolution.  

For example, direct imaging techniques such as coronagraphy or interferometry are affected by their Inner Working Angles (IWAs), which depend on the diameter of a telescope or the baseline of an interferometer, respectively.  For these instruments, the IWA defines a minimum angular scale such that their performance is degraded for angles smaller than the IWA.   For a coronagraph, the IWA scales as $IWA_C \sim3 \lambda/D$ where $\lambda$ is the wavelength of light, and $D$ is the telescope diameter.  For an interferometer, $IWA_I \sim \lambda / {2 B} $ where $B$ is the baseline of the interferometer.  An exoplanet at 1 AU around a Sun-like star  at 10 pc  is at an angular separation of 100 mas. This requires a telescope of diameter 3 m at 0.5 $\mu$m or an interferometer with a baseline of 10 m.   A thorough review of direct imaging of exoplanet techniques is presented in the review by Traub and Oppenheimer (2010), and community consensus reports on the state of the art of these techniques and future prospects are in Chapters 3 and 4 on optical and infrared imaging, respectively, in the Exoplanet Community Report edited by Lawson, Traub, and Unwin (2009).

 The present paper is built on our previous results, and it discusses PMS evolution as well as  evolution after the first ascent on the RGB including  the core-He burning phase all the way up to the Asymptotic Giant Branch (AGB).  We present these results for stars with solar metallicity having initial masses of 1.0, 1.5, and 2.0 $M_{\odot}$.  For the PMS evolution we use the tracks from Tognelli et al. (2011).\footnote{The Pisa PMS evolutionary tracks are available on the web at http://astro.df.unipi.it/stellar-models/.}  
  
For those same initial masses we also calculated the evolution of the habitable zone  over a wide range of metallicities. This time we started the evolution near the zero age main sequence.  As discussed above, we follow the evolution along the main sequence up to the RGB, then through the core He burning phase all the way to the tip of the  AGB.  We used a grid of five metallicities, Z=0.0001, 0.001, 0.017 (close to Solar), 0.040, and 0.070.  We used a helium abundance of Y=0.26 for all cases except for the highest metallicity, Z=0.070, where we used Y=0.30.    For our calculations we used the main sequence and post-main sequence tracks available from Girardi et al. (2000) and Bertelli et al. (2008).\footnote{The Padova tracks covering main-sequence and post-main sequence evolution up to the AGB are available on the web at http://stev.oapd.inaf.it/YZVAR.}

The Pisa and Padova tracks were chosen because the grid provided in the web sites covered the range of metallicities and masses of interest to this paper, and could be combined so that the luminosity evolution of a star from the PMS to the AGB could be consistently treated.  The tracks for Z=0.017 and Y=0.26, which we call ``near-solar metallicity'' in this paper, were the ``solar" abundance tracks  based on the historical standard solar composition (see the review by Grevesse and Sauval 1998).  In the past few years the initial metallicity of the Sun has been revised downwards to Z=0.015 (Lodders 2003).  The more recent value does not affect the results of this paper. 

In this paper we isolate the effect of metallicity on the evolution of the habitable zone by keeping the initial helium abundance, Y, a constant except at the highest metallicity value, Z=0.070, rather than using the helium enrichment law:  $Y(Z)  = Y_{p}+ (\Delta Y / \Delta Z) Z$, where $Y_p$ is the primordial helium abundance.  The helium enrichment law $(\Delta Y / \Delta Z)$ has been measured using eclipsing binaries (Ribas et al. 2000), giving  $(\Delta Y / \Delta Z)$ = 2.2 $\pm$ 0.8 with $Y_p$ = 0.225 $\pm$ 0.013, K dwarfs in the Hipparchos catalog with spectroscopically determined metallicities  (Jimenez et al. 2003) resulted in  $(\Delta Y / \Delta Z)$ = 2.1 $\pm$ 0.4, and field K dwarfs (Casagrande et al. 2007), giving $(\Delta Y / \Delta Z)$ = 2.1 $\pm$ 0.9.   If we adopt the most recent values for the  enrichment law $(\Delta Y / \Delta Z)$ = 2.1 $\pm$ 0.9, we find that  for Z=0.040, Y could be as low as  0.27 or as  high as 0.31.   These values are approximately within the spread of the measurements, as can be seen in Figure 7 of Ribas et al. (2000), which show  Z values near 0.015 have Y values between 0.20 and 0.29.  Given the large uncertainties in the helium enrichment law and the large spread in values of the initial helium abundance, it is reasonable to keep its value fixed except at the highest value of Z considered in this paper, where a larger value is clearly needed. 

The remainder of this paper is organized as follows.  Section 2 discusses the methodology employed in the computations and  Section 3 covers  the evolution of the habitable zone from the PMS to the AGB for 1.0, 1.5, and 2.0  $M_{\odot}$ stars.  Following this, in Section 4 we  describe our results on the effect of metallicity on the evolution of the habitable zone for a grid of five metallicities using a simple formulation for the equilibrium temperature of the exoplanet.  Next, in Section 5, we calculate the evolution of the habitable zone using a more refined estimate  for the inner and outer boundaries of the habitable zone, and we compare these results to the simpler formulation.  The connection between the initial metallicity parameter Z and the observable metallicity [Fe/H] is explored in Section 6.  In Section 7 we discuss the implications of our results on the search for life, and we present a summary of our findings in Section 8.  

\section{Evolution of Habitable Zone}

We follow the methodology of Paper I, in which the temperature of a planet at a chosen distance from its host star is computed as a function of time based on the luminosity evolution of the star.  As the star evolves and the luminosity changes, the inner and outer boundaries  of the habitable zone also change.  A planet is considered habitable if its equilibrium temperature is within the  range defined by the allowed temperature extremes.  In this paper we use an equilibrium temperature of 269 K for the inner boundary of the habitable zone and 169 K for the equilibrium temperature at the outer boundary of the habitable zone.   For the Sun, these extremes become 0.95 AU and 2.4 AU, respectively.    A more conservative limit for the equilibrium temperature of the outer boundary of the habitable zone is 203 K, which corresponds to 1.67 AU (Forget 1998).   These values assume the planet's Bond albedo is 0.2 as in Paper I. 

The  temperature of a planet is calculated based on the equilibrium between the amount of stellar radiation that is absorbed by the planet to the amount that is radiated by the planet back into space.  We use a greybody approximation in which the radiative properties of the planet are described by only a single parameter--the planet's albedo, $A$.  It is further assumed that the planet has perfect heat conductivity so that its temperature is uniform across its surface.  Thus, the planet's temperature, $T_p$ is:
\begin{equation}
\label{ }
T_p = {\left[ (1 - A) L_{\ast} / (16 \pi \sigma d^2) \right] }^{1/4} , 
\end{equation}
where $L_{\ast}$ is the luminosity of the star, $d$ the distance from the star to the planet, and $\sigma$ is the Stefan-Boltzmann constant.  For the calculations in this paper we assume a slightly higher Bond albedo of 0.3, which is Earth's average Bond albedo (de Pater \& Lissauer 2001).   Since the assumed albedo is a constant and thus wavelength independent, terms due to the albedo cancel out in the calculation.   

Unlike Paper I we consider only the less conservative limit for the temperature of the outer boundary of the habitable zone (169 K), which gives the longest possible durations of habitability.  The conservative limit for the temperature of the outer boundary of the habitable zone (203 K) was discussed in Paper I and it was based on Forget (1998). This temperature limit gives an outer habitable zone distance of 1.67 AU for a star like the Sun, and an albedo, $A=0.2$ (Paper I), or 1.58 AU for $A=0.3$ (this paper).  The outer limit is set by the lowest temperature at which the liquid-solid phase change of water can occur, which depends on the existence of a greenhouse effect from    
$\mathrm{CO_2 }$ and $\mathrm{{H_2}O} $  (Kasting et al. 1993, Kasting 1998).  
The less conservative temperature limit of 169 K, which corresponds to a distance of 2.4 AU ($A=0.2$) or 2.27 AU ($A=0.3$) for a star like the Sun, is based on planetary atmosphere models that  included $\mathrm{CO_2 }$ ice clouds.  Atmospheres with optically thick $\mathrm{CO_2 }$ ice clouds with large particle radii (6-8 $\mu$m) can maintain the surface of a planet above the freezing point of water (Forget \& Pierrehumbert 1997, Mischna et al. 2000).

Luminosity evolution is computed using the stellar evolution tracks provided by the Padova group (Bertelli et al.  2008) for the zero-age main sequence up to the first ascent along the RGB for a grid consisting of stars with $ M_{\ast}$ = 1.0, 1.5, and 2.0 $M_{\odot}$ and metallicities Z=0.0001, 0.001, 0.017, 0.040 with  helium abundance Y=0.26, and Z=0.070 with Y=0.030.  PMS evolution tracks were from the Pisa group (Tognelli et al. 2011) for Z=0.017 and Y=0.27, close to solar abundances and to those of the Padova group (Girardi et al. 2000, Bertelli et al. 2008), also for $M_{\ast}$ = 1.0, 1.5, and 2.0 $M_{\odot}$.  

In order to determine the starting time for the calculations of the transit of the habitable zone we compared the luminosity evolution from the Pisa tracks and the Padova tracks during the time periods in which they overlapped. The Pisa tracks started at the earliest PMS phase of core collapse at about $10^5$ years and ran to $\geqslant10^9$ years.  These tracks were compared to those from the Padova group.  The two sets of tracks produced luminosities within a few percent of each other over the time periods that overlapped.   Composite  tracks were generated from the PMS to the AGB by concatenating the two sets of tracks together were they closely matched.  We used the Pisa tracks from the PMS  up to 20, 40, and 80 million years for 1.0, 1.5, and 2.0 $M_{\odot}$, respectively.  We scaled the luminosities from these tracks to match the Padova tracks at those times, and generated a complete track by concatenating the two sets of tracks together into one file for each stellar mass selected.   The luminosities for these tracks also were within a few percent of the Pisa tracks, and a complete track was created in the manner described above.  Core He burning tracks from the Padova group were then concatenated with the composite tracks previously created.  

For the He flash we adopted a short and not significant duration compared to the habitability conditions we are evaluating of $10^5$ years, between the end of the first ascent along the RGB and the zero age of the core He burning phase.  The duration of the He flash adopted is consistent with the results of numerical simulations of the He flash performed by Dearborn, Lattanzio, and Eggleton (2006).  The computations were concluded at the tip of the AGB at the end of the core He burning tracks.   

 Figure 1 displays the resulting luminosity evolution  from $10^5$ years to $2 \times 10^{10}$ years for stars of 1.0 (solid line), 1.5 (dot-dashed line), and 2.0 (dashed line)  $M_{\odot}$, respectively.  For the 1 $M_{\odot}$ star case, note the decrease in luminosity during the core collapse, minimum luminosity at about $10^7$ years, and stable hydrogen burning starting at about 80 million years.   Note the large rise in luminosity over a short period of time for the RGB and AGB phases.  In this figure, the large asterisks represent the time when helium ignites in the core; the diamond symbols represent the approximate beginning of  the main sequence, when the star approaches hydrostatic equilibrium; and the triangles represent the time at which hydrogen is exhausted in the core.  The locations of the symbols displayed in the figure were taken directly from the evolutionary tracks.
 
 \begin{figure}
\epsscale{1.0}
\plotone{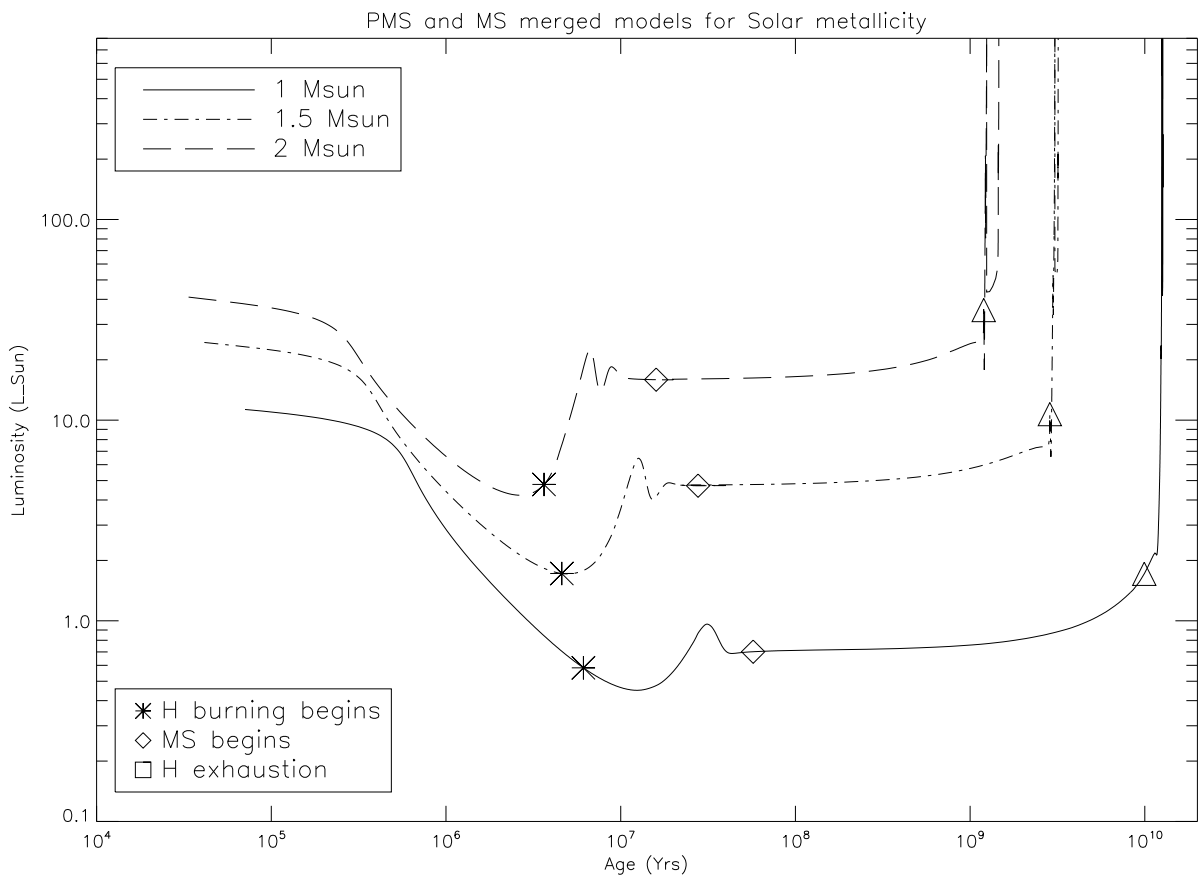}
\caption{Stellar luminosity as a function of the age of a star beginning in the PMS phase of core contraction and ending at the tip of the AGB for stars of 1.0 (solid line), 1.5 (dot-dashed line), and 2.0 (dashed line) Solar masses.  The large asterisk symbols represent the time when hydrogen ignites in the core; the diamond symbols represent the approximate beginning of  the main sequence, when the star approaches hydrostatic equilibrium; and the triangles represent the time at which hydrogen is exhausted in the core.  The values displayed in this figure were taken directly from the evolutionary tracks.}{\label{fig1}}
\end{figure}
 
 \section{PMS to AGB Evolution}

\begin{figure}
\epsscale{1.0}
\plotone{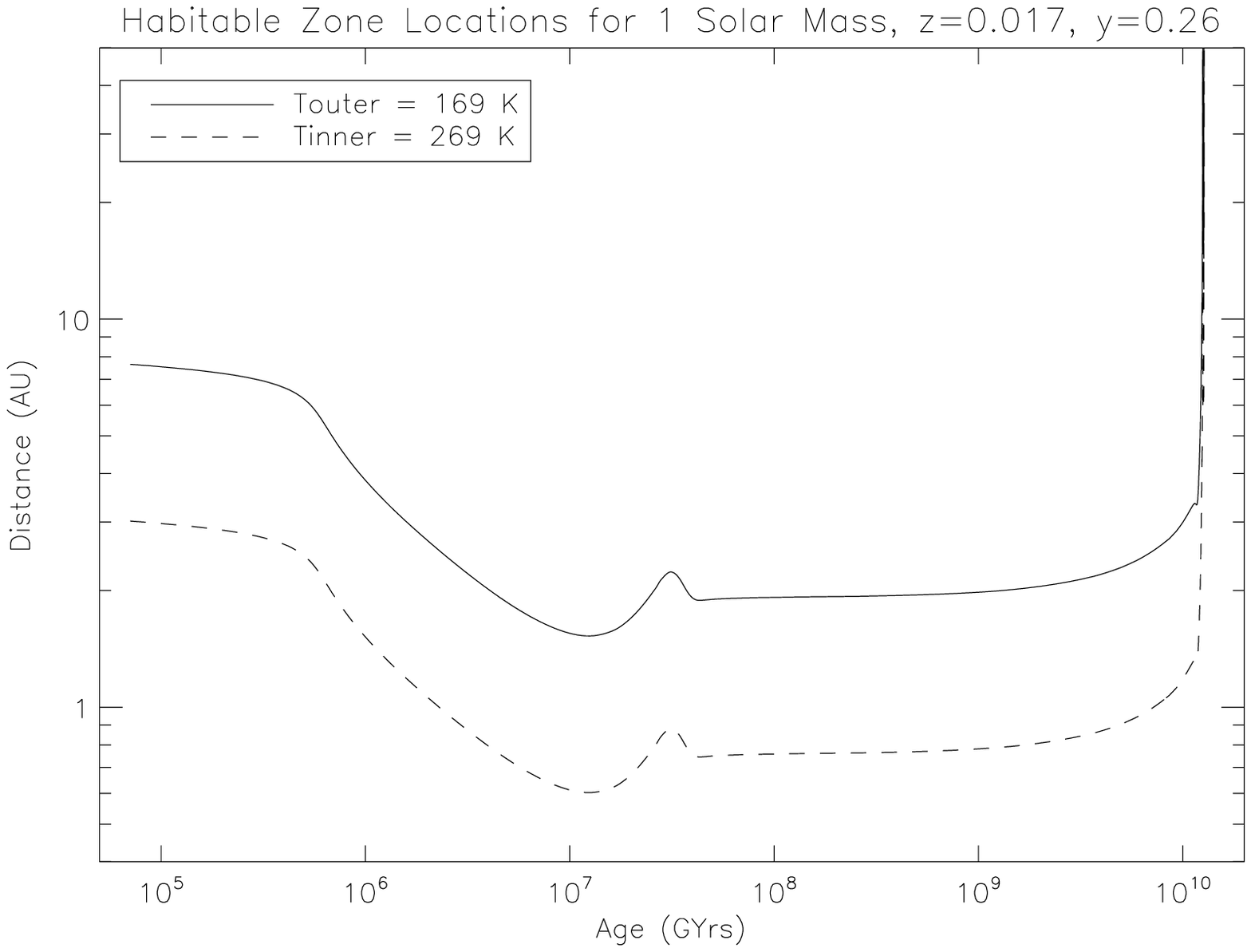}
\caption{Inner (dashed line) and outer (solid line) boundary locations for the habitable zone as a function of stellar age in (Gyr) for a 1.0 Solar mass star. }{\label{fig2}}
\end{figure}

Figure 2 displays the evolution of the inner (dashed line) and outer (solid line) boundaries of the habitable zone around a 1.0 $M_{\odot}$ star based on the luminosity evolution shown in Fig. 1.   Figure 3  displays the habitable zone boundary evolution on a linear scale, illustrating the very slow change during the first ten billion years and the very rapid rise in luminosity during the first ascent of the RGB.  Figure 4 is a close up of the last billion years of evolution highlighting the stable core helium burning phase from approximately 12.5 to 12.7  billion years.  Similarly, Figures 5 and 6 illustrate the evolution of the habitable zone for stars of 1.5 and 2.0 Solar masses, respectively.

\begin{figure}
\epsscale{1.0}
\plotone{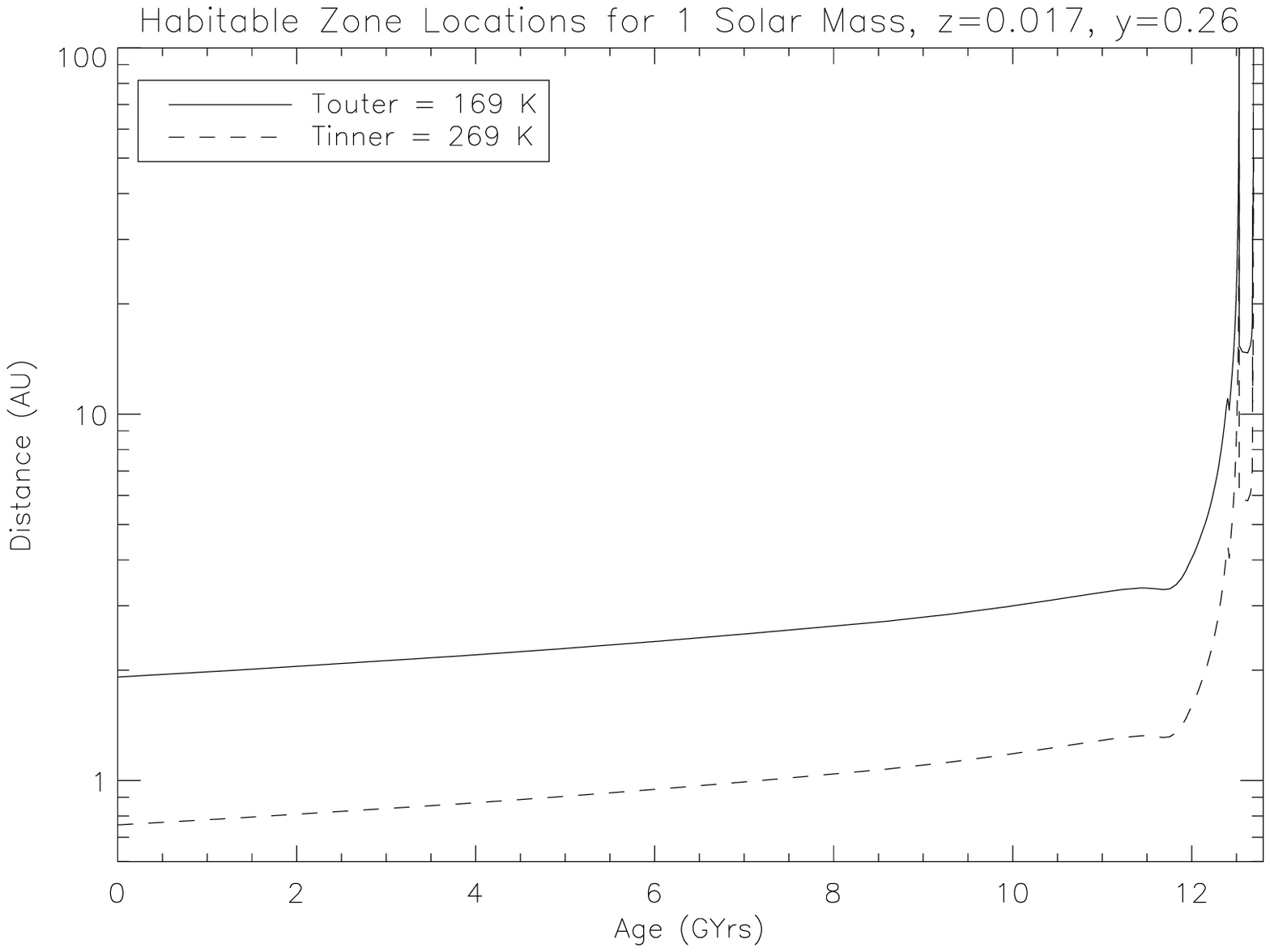}
\caption{Inner (dashed line) and outer (solid line)  boundary locations in AU for the habitable zone as a function of stellar age in Gyr for a 1.0 Solar mass star.}{\label{fig3}}
\end{figure}

\begin{figure}
\epsscale{1.0}
\plotone{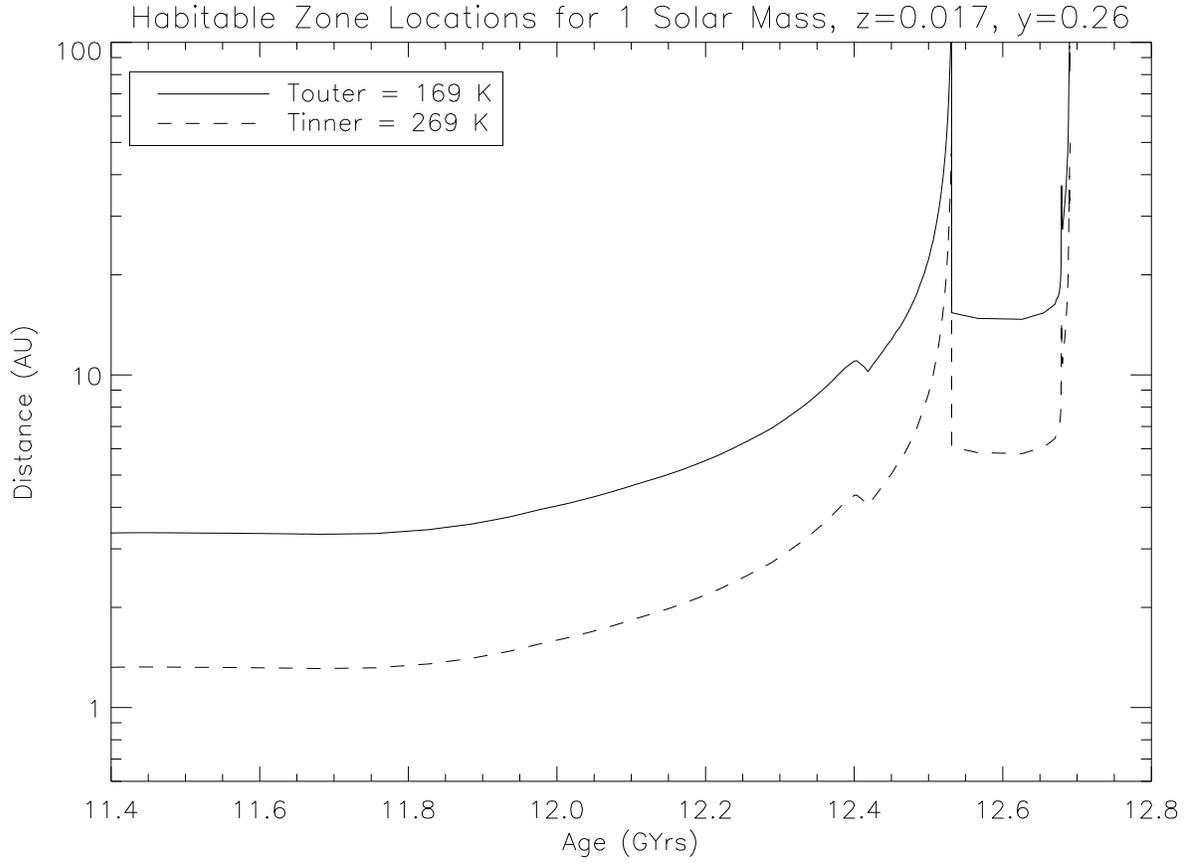}
\caption{ Inner (dashed line) and outer (solid line) boundary locations for the habitable zone as a function of stellar age for a 1.0 Solar mass star, but with a truncated scale to illustrate the evolution of the habitable zone during the core helium burning phase between 11.4 and 12.8 Gyr.}{\label{fig4}}
\end{figure}

\begin{figure}
\plotone{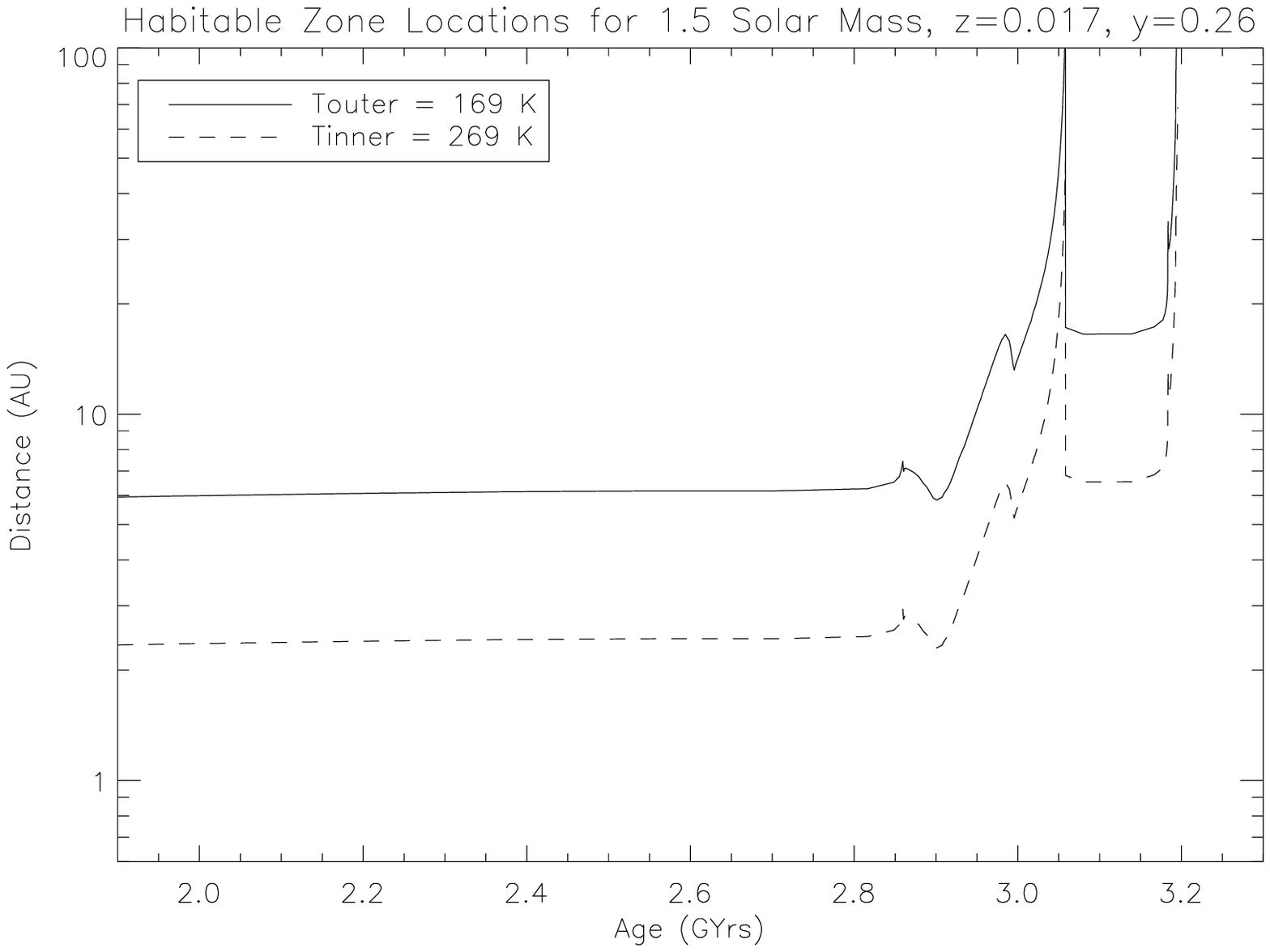}
\caption{Inner (dashed line) and outer (solid line)  boundary locations in AU for the habitable zone as a function of stellar age in Gyr for a 1.5 Solar mass star.}{\label{fig5}}
\end{figure}

\begin{figure}
\plotone{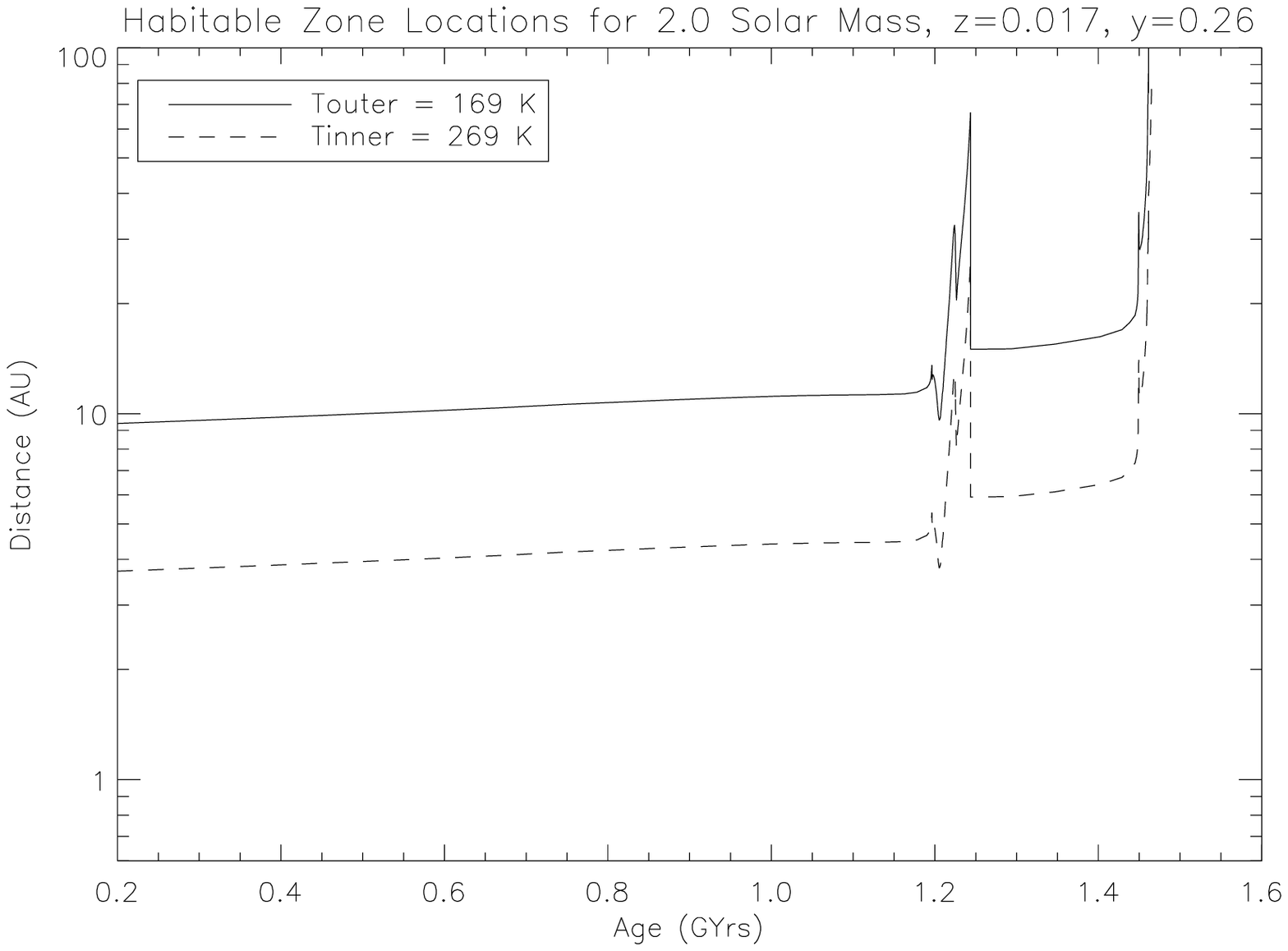}
\caption{Inner (dashed line) and outer (solid line)  boundary locations in AU for the habitable zone as a function of stellar age in Gyr for a 2.0 Solar mass star.}{\label{fig6}}
\end{figure}

The transit of the habitable zone at a fixed distance from the star or equivalently its duration is the time period of the luminosity evolution during which the equilibrium temperature of a planet calculated using Eq. (1) is between 269 K and 169 K.  In some cases there are multiple time periods during which a particular location is habitable.  In the figures we present only the longest duration if there is more than one time period.  This occurs at distances of greater than 10 AU from the star and is due to the rapid rise in luminosity along the first ascent of the RGB and the ascent up to the tip of the AGB.

\begin{figure}
\plotone{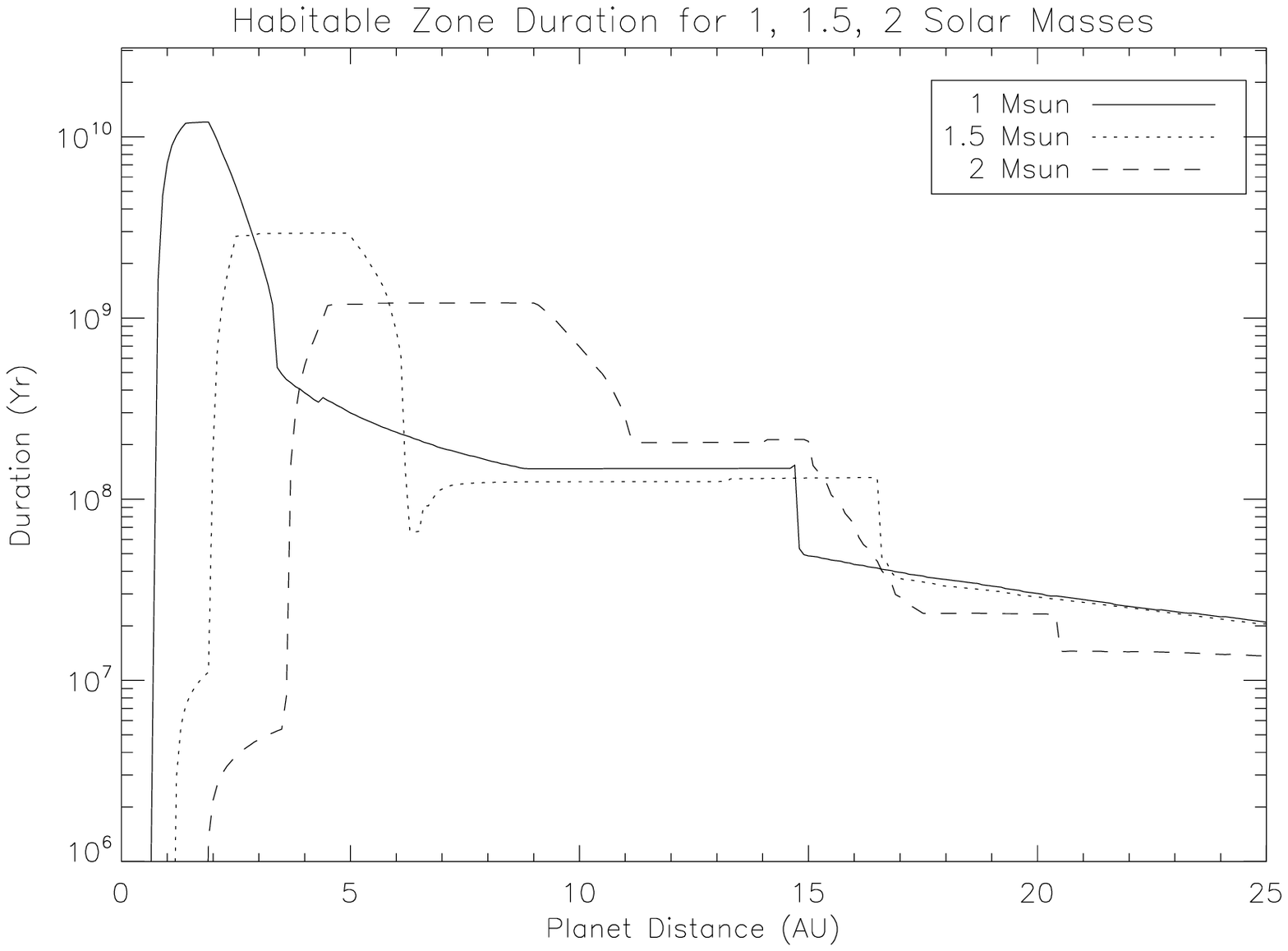}
\caption{Duration of transit of habitable zone for exoplanets around 1.0, 1.5, and 2.0 Solar mass stars as a function of distance from the star.  Near solar metallicity is assumed (Z=0.017). }{\label{fig7}}
\end{figure}

Figure 7 displays the duration of the transit of the habitable zone for 1.0 (solid line), 1.5 (dotted line), and 2.0 (dashed line) $M_{\odot}$ stars as a function of the distance of an exoplanet from its host star.  For 1.0 $M_{\odot}$ stars, the longest duration is strongly peaked between 1 and 2 AU, for 1.5 $M_{\odot}$, the peak is broader and further away from the star, i.e., from 2 to 5 AU.  For 2.0 $M_{\odot}$ stars the peak is still broader and is from 4 to 9 AU.   The increasing width of the peak of the duration of the habitable zone  as the stellar mass  increases is as expected due to the increase in  luminosity during the main sequence as the mass of the star increases.  

The duration of the transit of the habitable zone for each stellar mass shown in  Fig. 7 are in agreement with Paper I.  However, in this paper, we include the period during core He burning, which produces  a broad (in width in AU) and flat duration of $\sim10^8$ years, for each stellar mass chosen.  This gives a modest chance for life again after the He flash for exoplanets at distances  of 8-15 AU for a 1 $M_{\odot}$ star, 6-16 AU for a 1.5 $M_{\odot}$ star, and 12-15 AU for a 2.0 $M_{\odot}$ star.    The reason that the duration is essentially independent of the initial stellar mass is because the evolutionary models predict that the mass of the He in the core is approximately the same during quiescent core He burning regardless of the initial mass.  Thus the lifetime of the core He burning phase is approximately independent of the initial mass, which means the duration of the transit is approximately the same.  

This result is somewhat different than was discussed in Paper I, in which we used the empirical formula of Scalo and Miller (1979) to estimate the lifetime of the core He burning phase.  Our previous estimate was as high as $10^9$ years for the core He burning phase for a 1  $M_{\odot}$ star using this formula.  The models used in this paper all give lifetimes of the core He burning phase of about $2 \times 10^8$ years, somewhat lower than in Paper I.

\section{Effect of Metallicity on the Evolution of the Habitable Zone}

\begin{figure}
\plotone{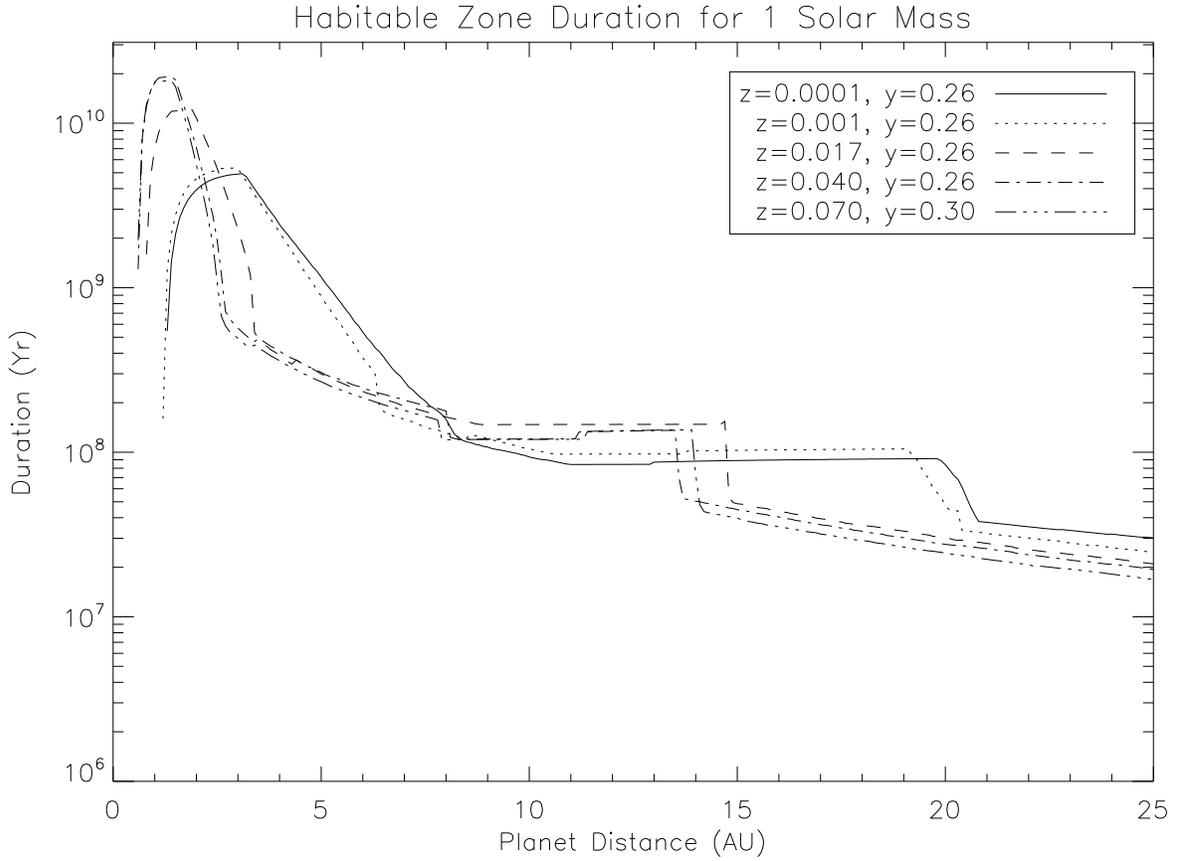}
\caption{Duration of transit of habitable zone for exoplanets around a 1.0 Solar mass star as a function of distance from the star.  In this figure a grid of metallicities is assumed (Z=0.0001, 0.001, 0.017, 0.040, 0.070).  The Solar value for the He abundance, Y=0.26, is assumed for all curves except for Z=0.070, where Y=0.30 is assumed.  }{\label{fig8}}
\end{figure}

Metallicity has a substantial effect on the duration of the transit of the habitable zone.  Figure 8 displays the duration of habitability as a function of exoplanet distance for a grid of five metallicities (Z=0.0001, 0.001, 0.017, 0.040, 0.070) for a star of 1 $M_{\odot}$.  Both the location and the width of the longest duration change substantially as a function of metallicity.  The longest duration and narrowest width is for the high metallicity cases, Z=0.040 (dot-dashed line) and Z=0.070 (dot-dot-dot-dashed line) in which the duration is strongly peaked right at slightly more than 1 AU.  Durations as long as $2 \times 10^{10}$ are possible.  The width in AU (important for observational considerations) for durations longer than $10^9$ years is about 2 AU.  For near Solar metallicity stars (Z=0.017, dashed line), as in Fig. 7, the duration peaks at slightly more than $10^{10}$ years and has a width of about 2 AU with a peak at about 1.7 AU.   Low metallicity stars with Z=0.001 (dotted line) and Z=0.0001 (solid line) have a peak duration of about $4 \times 10^9$ years at 3 AU with a width of 4 AU.  

\begin{figure}
\plotone{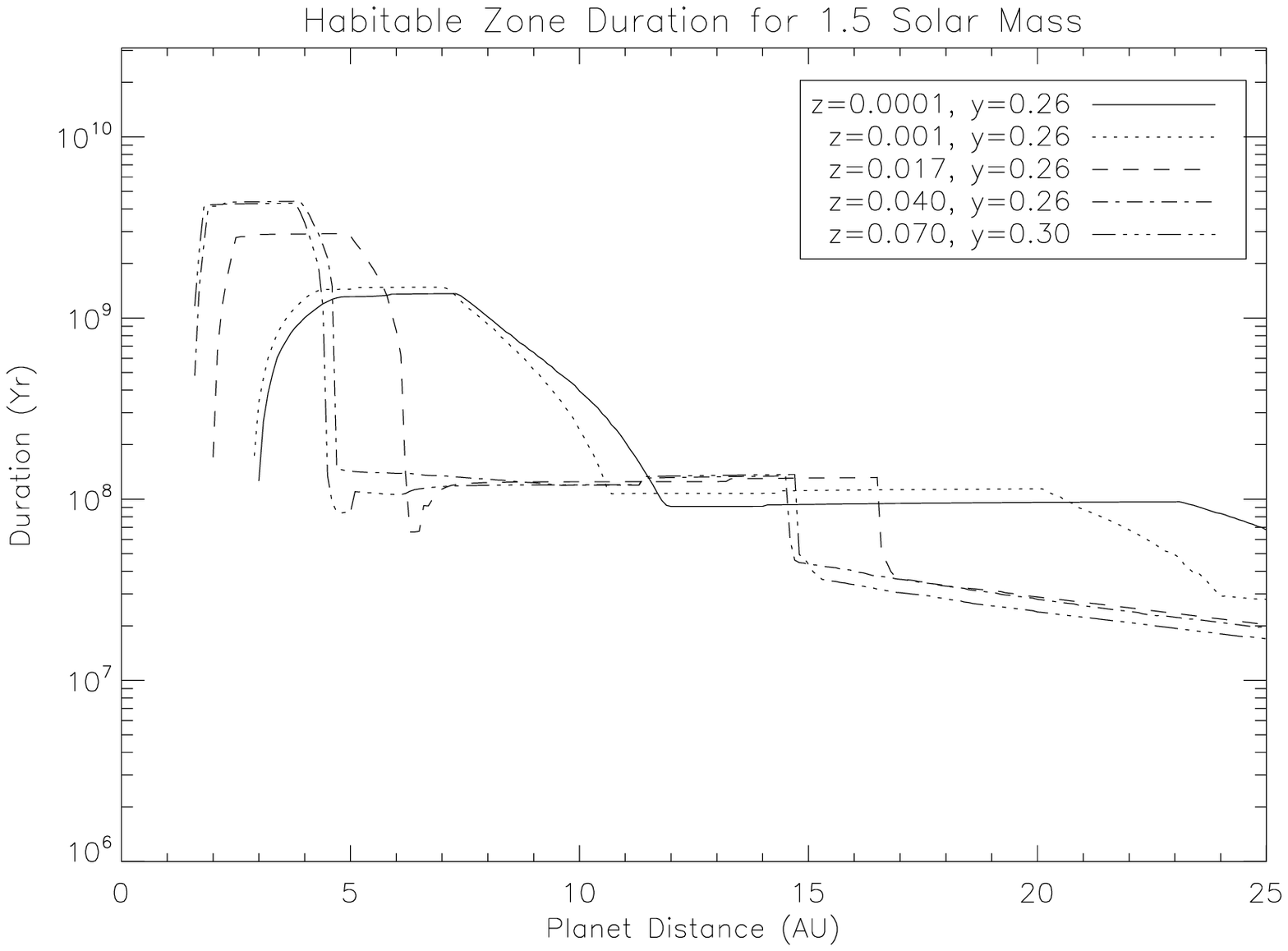}
\caption{ Duration of transit of habitable zone for exoplanets around a 1.5 Solar mass star as a function of distance from the star.  In this figure a grid of metallicities is assumed (Z=0.0001, 0.001, 0.017, 0.040, 0.070).  The Solar value for the helium abundance, Y=0.26, is assumed for all curves except for Z=0.070, where Y=0.30 is assumed. }{\label{fig9}}
\end{figure}

Similar behavior, i.e., shortening of the habitable zone duration, is seen as stellar mass increases.  Figure 9 shows the duration of the transit of the habitable zone for a 1.5 $M_{\odot}$ star for the same grid of metallicities used in Fig. 8.  In this case the duration of the habitable zone for exoplanets around high metallicity stars is reduced (compared to a high metallicity 1 $M_{\odot}$ star) to a maximum of $3 \times 10^9$ years centered at 3 AU with a width of 2.5 AU.  The duration is reduced as well for exoplanets around Solar metallicity stars to $2 \times 10^9$ years and a width of 4 AU.  The duration is reduced still further to just over $10^9$ years for the high metallicity case, with a width of about 5 AU.  As in Fig. 8, there is a broad region from about 5 to 15 AU during the core He burning phase for Solar and high metallicity stars that lasts somewhat longer than $10^8$ years.  For low metallicity stars this region is from about 12 to 24 AU.  

\begin{figure}
\plotone{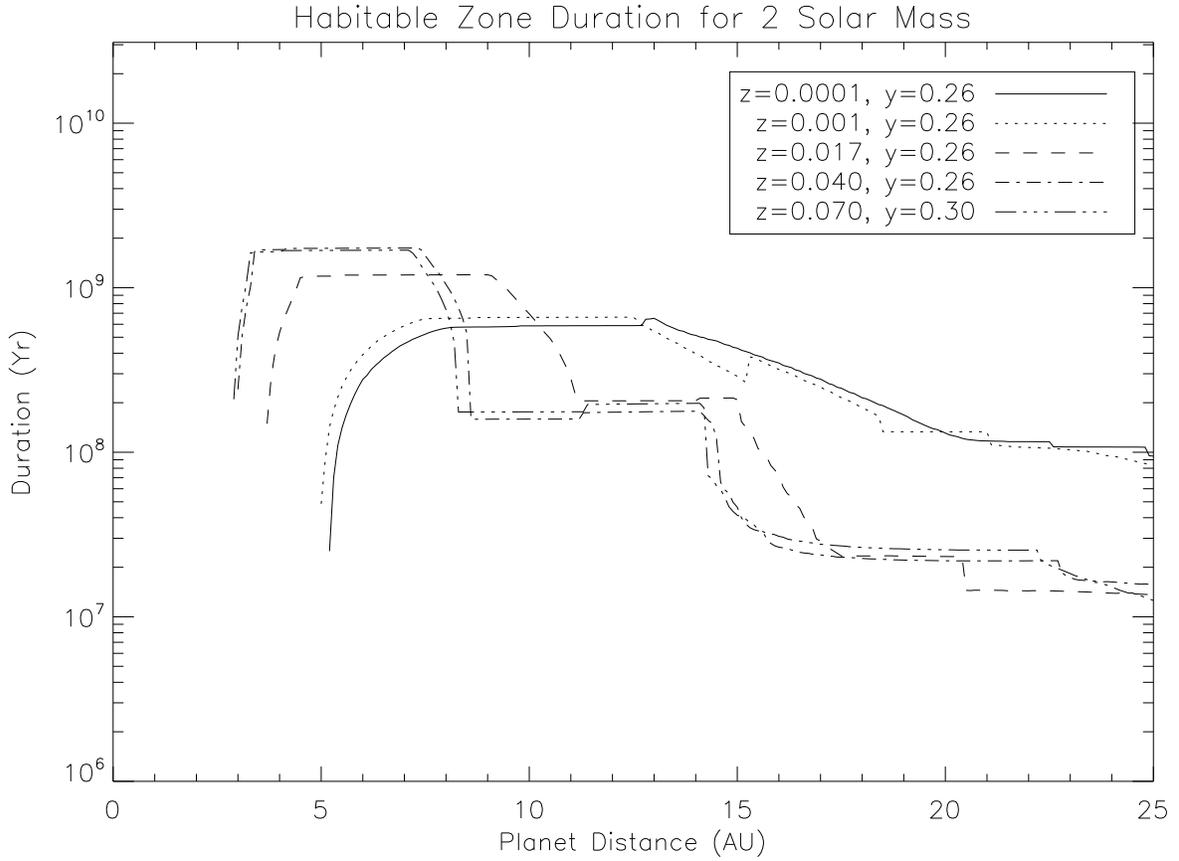}
\caption{Duration of transit of habitable zone for exoplanets around a 2.0 Solar mass star as a function of distance from the star.  In this figure a grid of metallicities is assumed (Z=0.0001, 0.001, 0.017, 0.040, 0.070).  The Solar value for the He abundance, Y=0.26, is assumed for all curves except for Z=0.070, where Y=0.30 is assumed. }{\label{fig10}}
\end{figure}

Figure 10 shows the duration of the transit of the habitable zone for a 2.0 $M_{\odot}$ star for the same grid of metallicities used in Fig. 9.  In this case the duration of the habitable zone for exoplanets around for high metallicity stars is reduced to a maximum of $ \lesssim 2 \times 10^9$ years centered at $\sim$ 5 AU with a width of $\sim$ 4 AU.  The duration is reduced somewhat for exoplanets around Solar metallicity stars to $ \sim 1 \times 10^9$ years and a width of $\sim$ 5 AU.  The duration is reduced substantially to just over $\sim 6 \times 10^8$ years for the high metallicity case, with a width of about 6 AU.  As in Fig. 9, there is a broad region from about 8 to 14 AU during the core He burning phase for Solar and high metallicity stars that lasts somewhat longer than $10^8$ years.  For low metallicity stars this region is from about 20 to $\gtrsim$ 25 AU.  

\section{Effect of Improved Estimates of Inner and Outer Boundaries of the Habitable Zone on the Duration of Habitability}

In this paper we focused on how the metallicity of the host star affects the location and duration of the habitable zone.  We have used a very simple formulation to determine the inner and outer limits of the habitable zone by considering only the effective temperature of the planet as determined by radiative equilibrium based only on the luminosity of the host star.  Recently, the choice of equilibrium temperatures for the habitable zone has been revisited in the context of potentially habitable transiting planets discovered by Kepler (Borucki et al. 2011, Kaltenegger \& Sasselov 2011).  A more complex formula for the location of the inner and outer radius of the habitable zone is discussed in Kaltenegger et al. (2011), based on the equations of Selsis et al. (2007).   These relations for the inner and outer boundaries of the habitable zone take into account the radiative properties of the atmosphere of the planet by adding correction factors based on the difference in the effective temperature of the star compared to that of the Sun.  The formulae  are:  

\begin{equation}
\label{ }
l_{in} = (l_{in\_sun} - a_{in} T_{star} - b_{in} T_{star}^2) (L/L_{sun})^{1/2}
\end{equation}

\begin{equation}
\label{ }
l_{out} = (l_{out\_sun} - a_{out} T_{star} - b_{out} T_{star}^2) (L/L_{sun})^{1/2}
\end{equation}

In these equations, $T_{star} = T_{eff} - 5700$, $a_{in} = 2.7619 \times 10^{-5}$, $b_{in} = 3.8095 \times 10^{-9}$,
$a_{out} = 1.3786 \times 10^{-4}$, $b_{out} = 1.4286 \times 10^{-9}$, and the inner and outer radii are $l_{in}$ and 
$l_{out}$ in AU, respectively, and $T_{eff}$ is in K.   These equations are valid from 3700 K to 7200 K.  Figure 13 displays the effect of this correction to the inner and outer locations of the habitable zone, as compared to the situation around the Sun.  Here we fix L to be equal to Solar luminosity, and we derive the fractional change in the inner and outer radii of the habitable zone, based on $l_{in\_sun} = $ 0.90 AU, and $l_{out\_sun} =$ 2.27 AU  using the albedo A=0.3 used throughout this paper.  With this scaling we isolate the effect of the terms in the equations depending on the parameter 
$T_{star}$.   The upper panel displays the fractional correction to the inner (solid line) and outer (dashed line) boundaries of the habitable zone.  For the inner boundary, the corrections are about $+$4\% at 3700 K and $-$5\% at 7200 K, while the outer boundary corrections are at the $+$10\% and $-$9\% levels for 3700 K and 7200 K, respectively.  Based on this analysis the width of the habitable zone is also affected.  The bottom panel of Figure 13 displays the fractional correction to the width of the habitable zone as a function of the effective temperature of the star.  Note that the effect is at the $+$15\% level for the lowest effective temperature and $-$9\% at the highest effective temperature considered. 

\begin{figure}
\epsscale{0.85}
\plotone{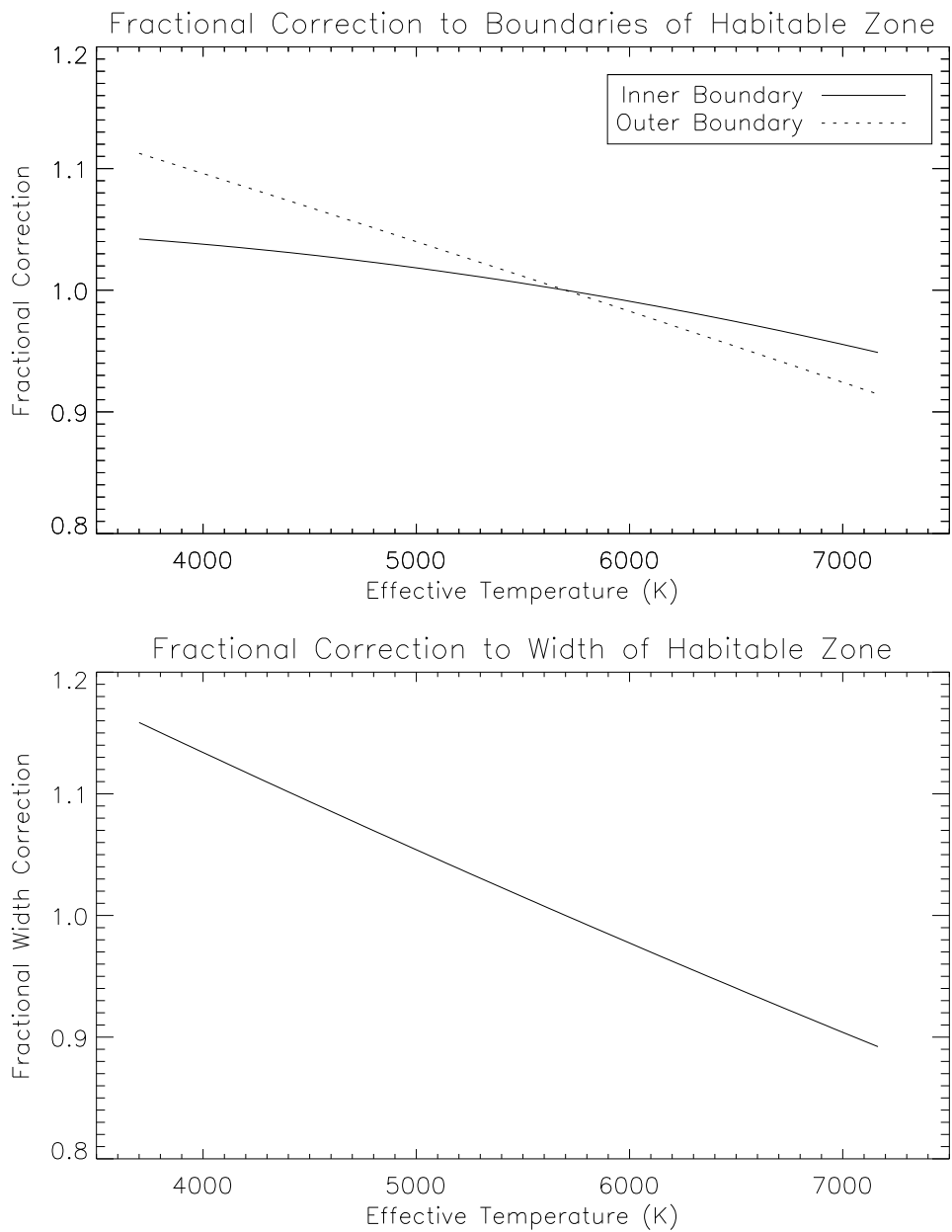}
\caption{Top panel:  Fractional correction to inner (solid line) and outer (dotted line) location of the habitable zone as a function of stellar effective temperature.   The correction to the habitable zone inner and outer boundaries are due to the change in spectrum of light from the star which is either bluer or redder depending on its evolutionary state and its initial mass.  The amount of radiation absorbed or trapped in the atmosphere of the planet is affected by the stellar spectrum.  Bottom panel:  Fractional change in width of habitable zone  as a function of effective  temperature based on corrections in top panel.  These corrections are normalized to the effective temperature of the Sun and the inner radius of 0.90 AU and outer radius of 2.27 AU for the habitable zone for the Solar environment.}{\label{fig11}}
\end{figure}

These improved estimates of the inner and outer boundaries of the habitable zone may have implications for the duration of the habitable zone.  In the remainder of this section we explore their applicability and implications in terms of the duration of the habitable zone.  

\begin{figure}
\epsscale{1.0}
\plotone{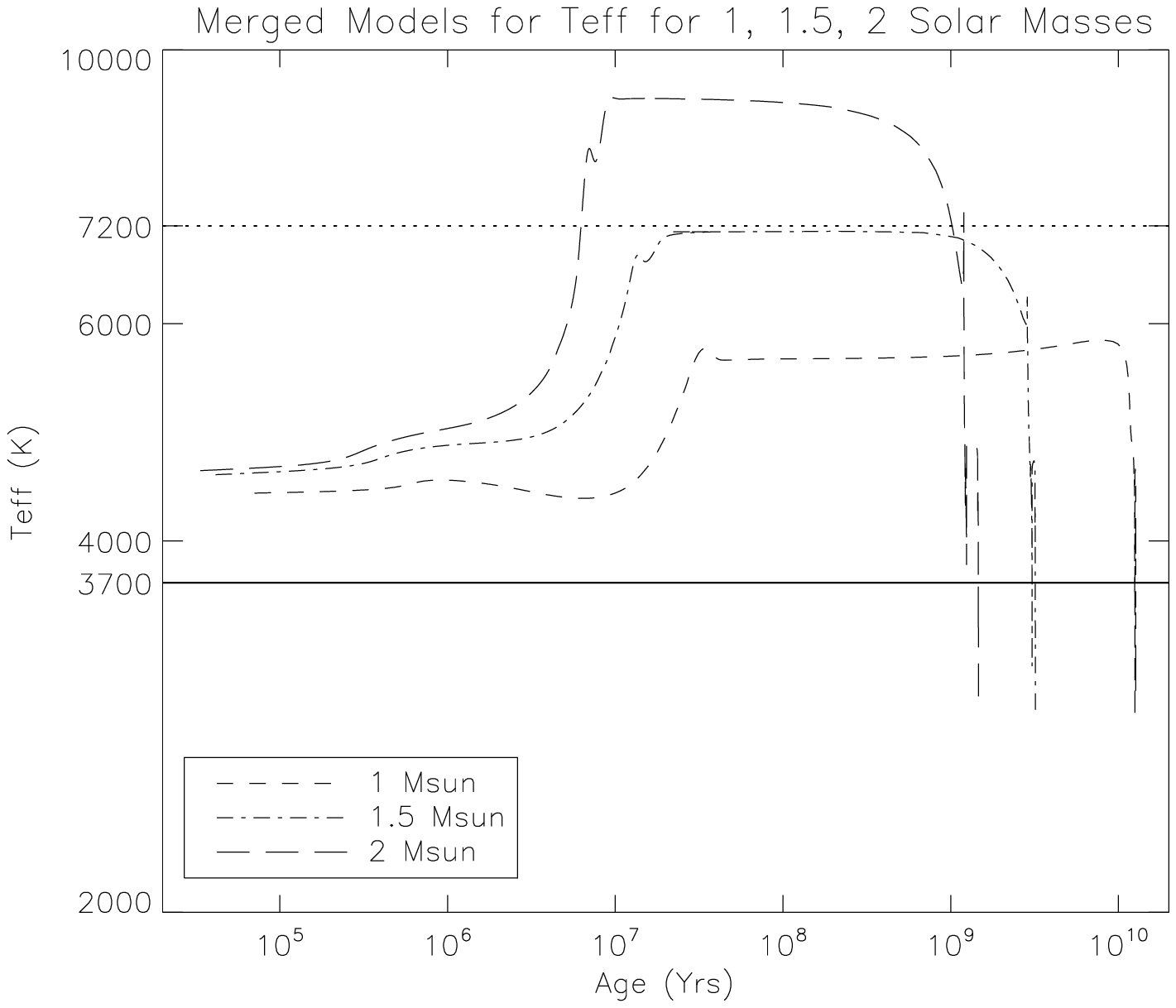}
\caption{Effective temperature of 1.0, 1.5, and 2.0 $M_{\odot}$ stars as a function of age.  Note that the formulae for the improved estimate of the inner and outer boundaries of the habitable zone correspond to 3700 K (solid line) and 7200 K (dotted line).  Thus, the formulae are not applicable to 2.0 $M_{\odot}$ stars and are applicable to 1.5 $M_{\odot}$ stars for solar and higher metallicities.}{\label{fig12}}
\end{figure}

Figure 12 displays the effective temperature as a function of stellar age for 1.0, 1.5 and 2.0 $M_{\odot}$ stars for near solar metallicity values Z=0.017, Y=0.26 as used in the previous section.  As previously discussed, Eqs. (2) and (3) can be used for stars with effective temperatures from 3700 K to 7200 K, which are shown as a solid line and a dashed line, respectively, in the figure.  We find that  these equations are applicable for 1 $M_{\odot}$ stars  for all the metallicities considered and for 1.5 $M_{\odot}$ stars for Solar and higher metallicities.   Note that the effective temperature of the  1.5 $M_{\odot}$ star is barely outside of the limits of the equations, so we apply them to this case for completeness.  

\begin{figure}
\epsscale{1.0}
\plotone{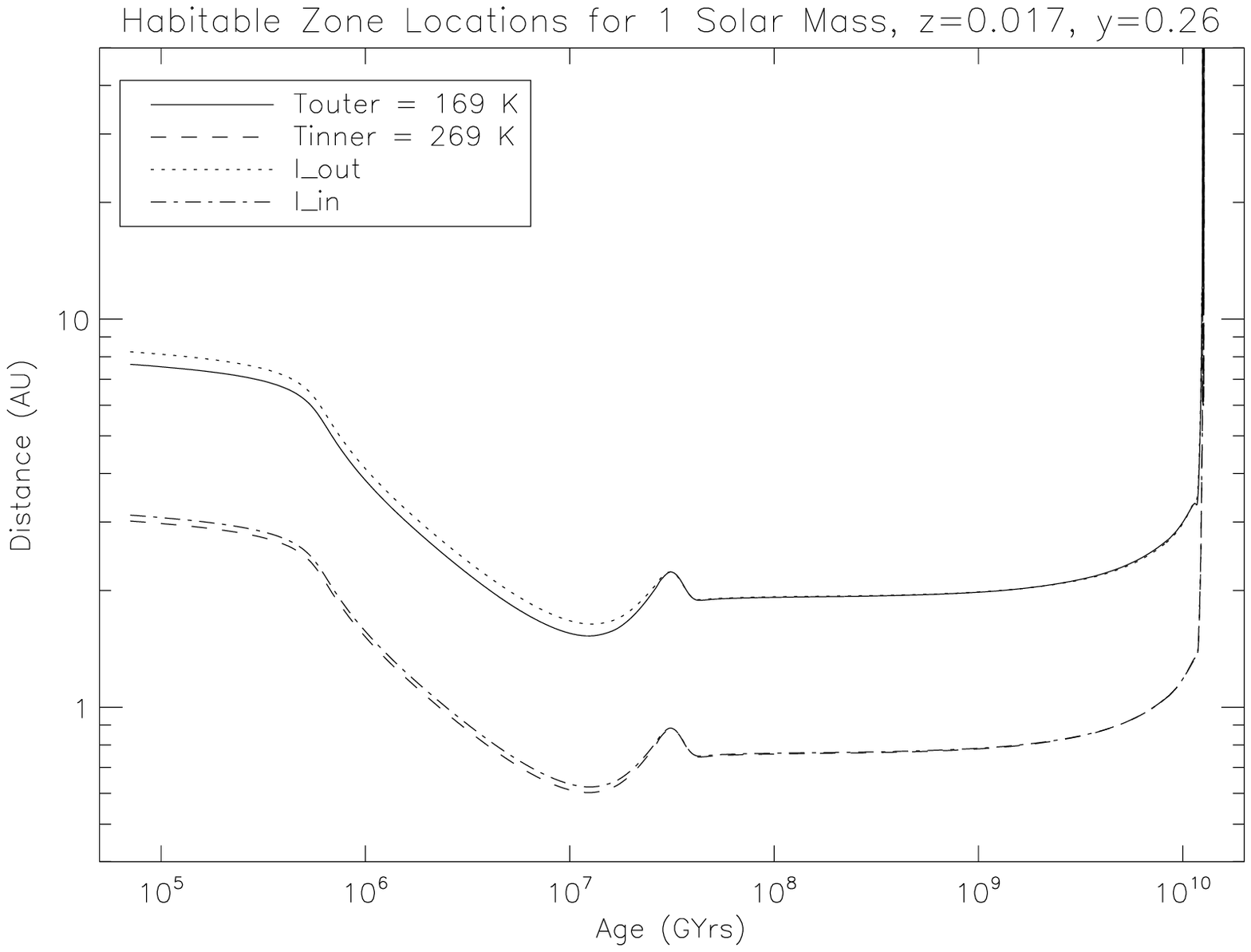}
\caption{Inner (dashed line) and outer (solid line) boundary locations for the habitable zone as a function of stellar age in (Gyr) for a 1.0 Solar mass star assuming the simple model for the habitable zone boundaries, Eq. (1) in the text.  Dot-dashed and dotted lines represent the inner and outer boundaries of the habitable zone for the more refined model, using Eqs. (2) and (3) in the text. }{\label{fig13}}
\end{figure}

The main implication of these equations is that $l_{in}$ and $l_{out}$ give slightly different values for the inner and outer boundaries of the habitable zone than using an equilibrium temperature for the planet based only on the luminosity of the star.  Figure 13 displays the results for a 1 $M_{\odot}$ star.   During the PMS phase, the inner boundary of the habitable zone is about a hundredth of an AU further than for the simpler estimate, and the outer boundary is almost an AU further out, i.e., from nearly an AU at $10^5$ years and down to a fraction of an AU at $10^7$ years.  It is clear that there is no significant difference in the location of the habitable zone during the main sequence as the curves lie on top of each other.  Figure 14 displays the effect of the improved formulae for the late main sequence and post main sequence phases.  We see that as in the PMS phases, the locations of the inner and outer boundaries move outward only fractions of an AU. 

\begin{figure}
\epsscale{1.0}
\plotone{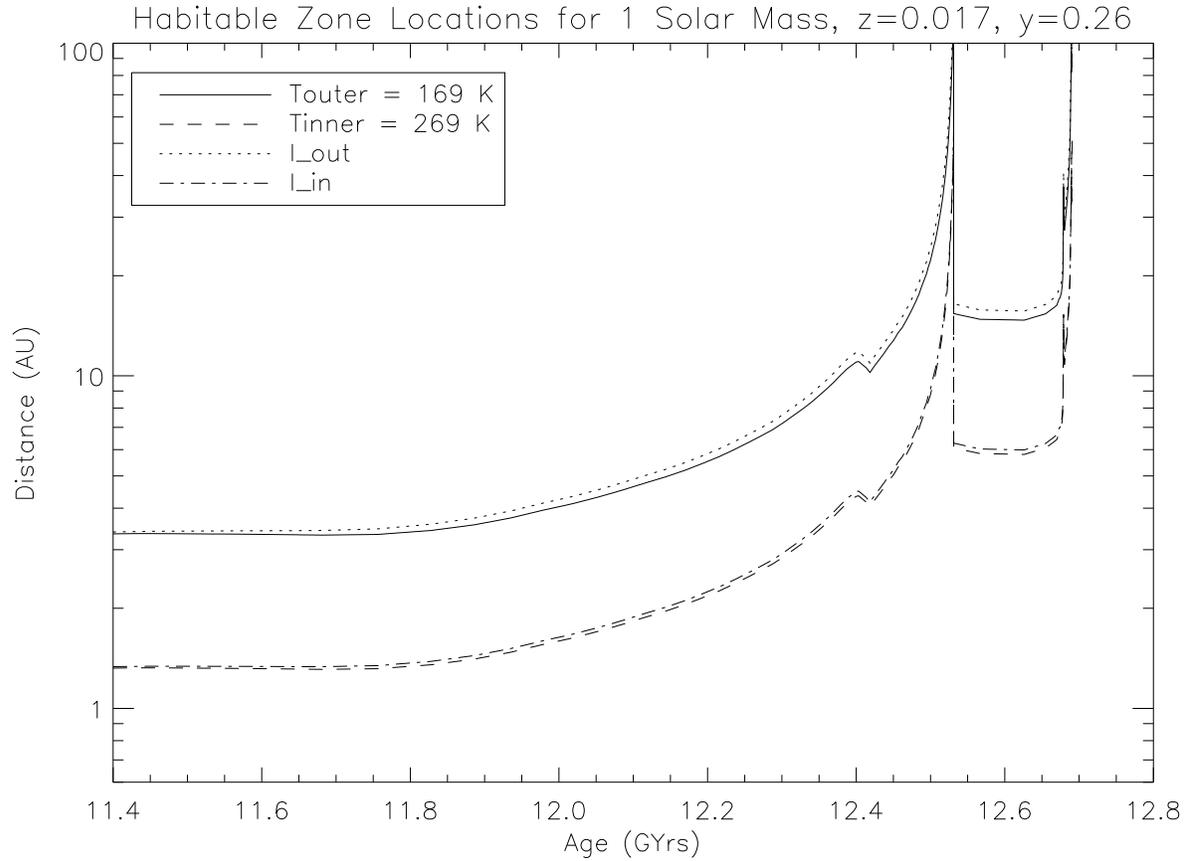}
\caption{ Inner (dashed line) and outer (solid line) boundary locations for the habitable zone as a function of stellar age for a 1 Solar mass star, but with a truncated scale to illustrate the evolution of the habitable zone during the core helium burning phase between 11.4 and 12.8 Gyr.  Dot-dashed and dotted lines represent the inner and outer boundaries of the habitable zone for the more refined model.}{\label{fig14}}
\end{figure}

\begin{figure}
\plotone{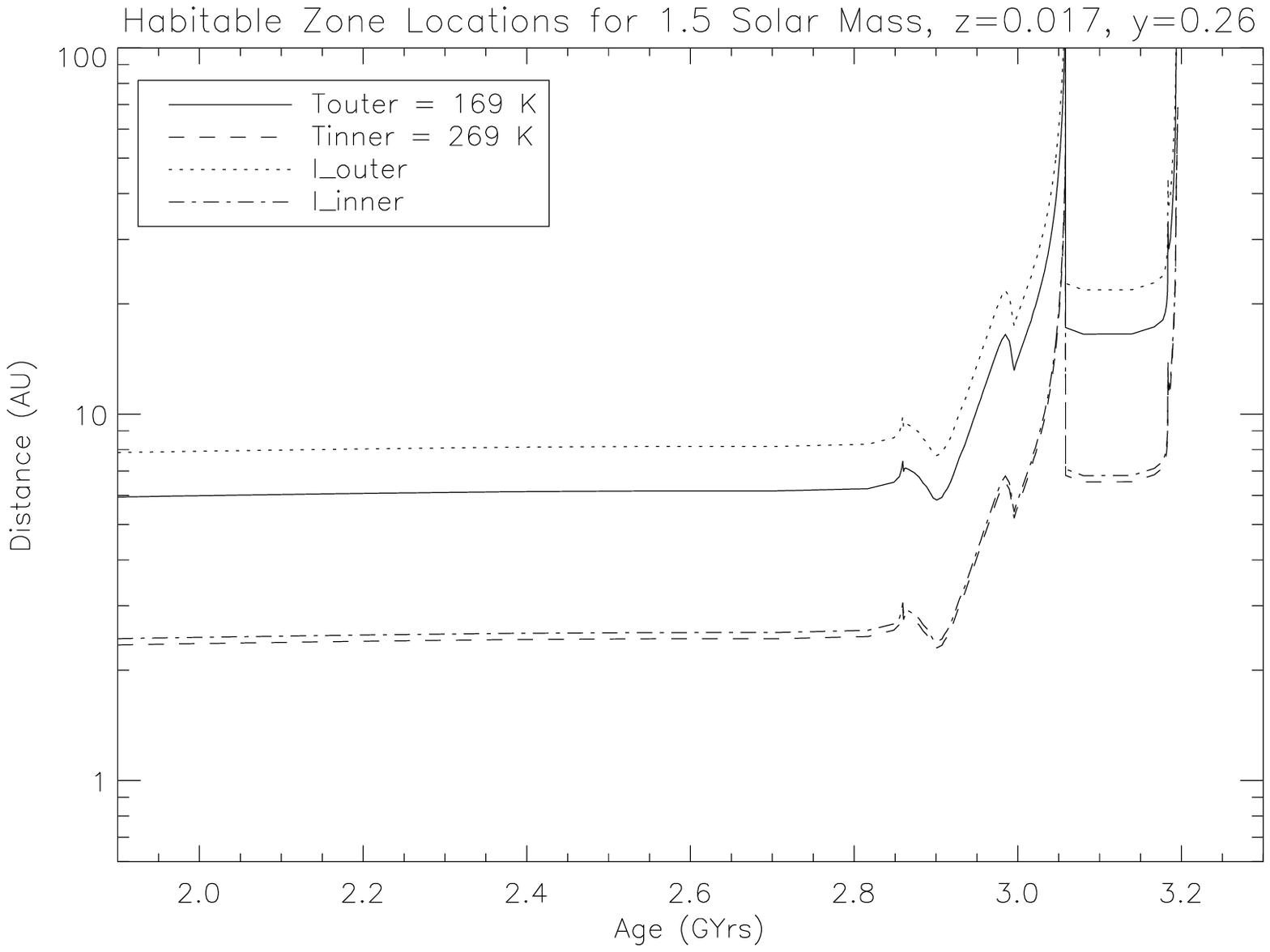}
\caption{Inner (dashed line) and outer (solid line)  boundary locations in AU for the habitable zone as a function of stellar age in Gyr for a 1.5 Solar mass star.  Dot-dashed and dotted lines represent the inner and outer boundaries of the habitable zone for the more refined model.
}{\label{fig15}}
\end{figure}

For a 1.5 $M_{\odot}$ star, the movement of the outer boundary is more significant as can be seen in Fig. 15.   At about 2 Gyr the outer boundary is 6 AU for the case with $T_{outer}$ = 169 K compared with $l_{outer}$ which is at 8 AU, giving a difference of about 2 AU, with a similar change during the core He burning phase at about 3.1 Gyr. The inner boundary change is small, however, amounting to a fraction of an AU.  

\begin{figure}
\plotone{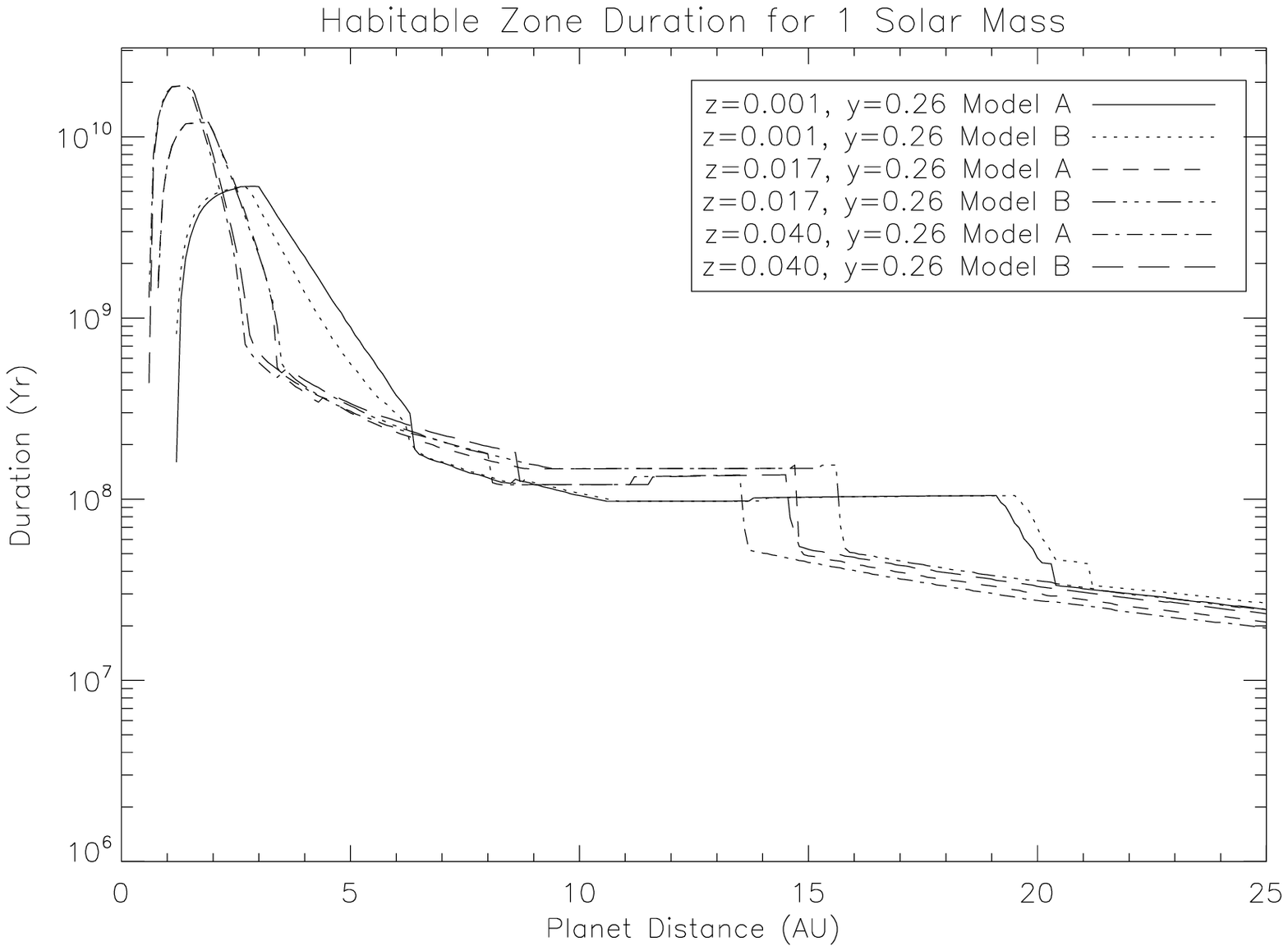}
\caption{Duration of transit of habitable zone for exoplanets around a 1.0 Solar mass star as a function of distance from the star.  In this figure a grid of metallicities is assumed (Z=0.0001, 0.001, 0.017, 0.040, 0.070).  The Solar value for the Helium abundance, Y=0.26, is assumed for all curves except for Z=0.070, where Y=0.30 is assumed. Model A in the legend refers to the simple formulation for the boundaries of the habitable zone, Eq. (1) in the text, and Model B represents the more refined formulae of Eqs. (2) and (3).   
}{\label{fig16}}
\end{figure}

We can now calculate how the two formulae for the inner and outer boundaries of the habitable zone affect the duration of habitability at a particular distance from the star.  Figure 16 displays the duration of habitability as a function of distance from the star for three different metallicities (Z=0.001, 0.017, and 0.040) for the two estimates of the inner and outer boundaries of the habitable zone.  In this figure, the curves labeled  ``Model A''  represent the duration based on Eq. (1), while the curves labeled ``Model B'' are based on Eqs. (2) and (3).  It is clear that neither the maximum duration nor the width is strongly affected for Z=0.017 and Z=0.040, i.e., solar and higher metallicities.  However, there is a slight decrease in the width of the curve for the low metallicity case, Z=0.001, with the curve from Model B having a slightly narrower peak, amounting to $\sim$ 0.5 AU for durations greater than about 1 Gyr.  The effect is opposite in sign at the later stages of stellar evolution, with the widths from Model B about 0.5 to 1 AU wider than the corresponding results from Model A.  

\begin{figure}
\plotone{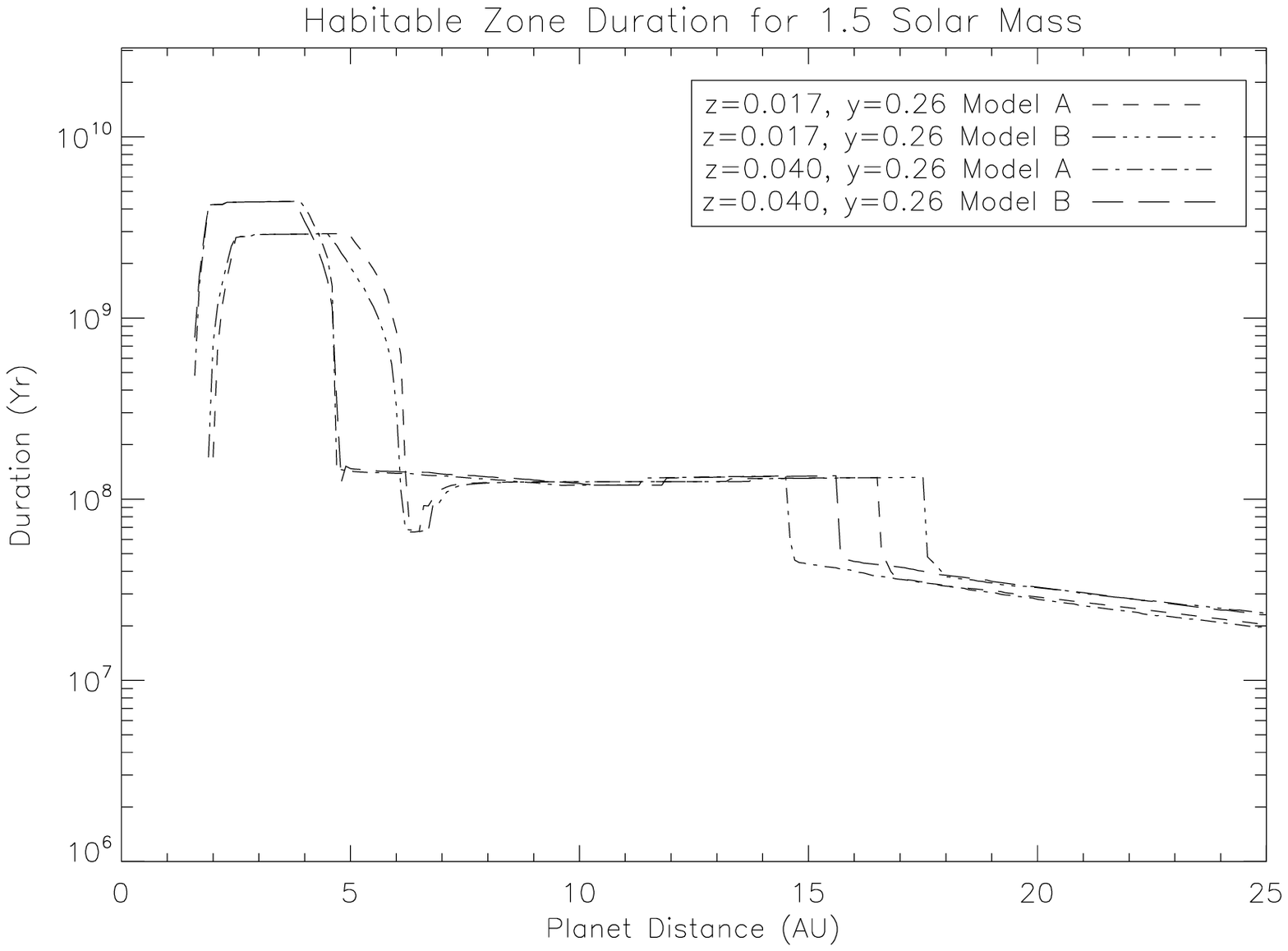}
\caption{ Duration of transit of habitable zone for exoplanets around a 1.5 Solar mass star as a function of distance from the star.  In this figure a grid of metallicities is assumed (Z=0.0001, 0.001, 0.017, 0.040, 0.070).  The Solar value for the He abundance, Y=0.26, is assumed for all curves except for Z=0.070, where Y=0.30 is assumed.  Model A in the legend refers to the simple formulation for the boundaries of the habitable zone, Eq. (1) in the text, and Model B represents the more refined formulation.}{\label{fig17}}
\end{figure}

Similar results have been obtained for a 1.5 $M_{\odot}$ star as can be seen in Fig. 17.  The difference in widths at the peak duration for Z=0.017 amounts to about 0.2 AU, and is about 0.5 AU for the higher metallicity case (Z=0.040), with Model A being slightly wider.    As in Fig. 16, the sign of the effect reverses and Model B is about 1 AU wider for the two metallicities. 

\begin{table}
\begin{center}
\caption{[Fe/H] values from Z\label{tbl-3}}
\begin{tabular}{ccccc}
\tableline \tableline
Z & [Fe/H]\tablenotemark{a} & [Fe/H]\tablenotemark{b}&  [Fe/H]\tablenotemark{c} & [Fe/H]\tablenotemark{d}   \\
\tableline
0.0001 & -2.09 & -2.09 & -2.11  & -2.11 \\
0.001 & -1.09 & -1.08 & -1.11 & -1.10 \\
0.017 & 0.15  & 0.18 & 0.13 & 0.14  \\
0.040 &0.54 & 0.60 & 0.52  & 0.55 \\
0.070 & 0.83 & 0.95 & 0.81 & 0.87 \\
\tableline
\end{tabular}
\tablenotetext{a}{[Fe/H] calculated using values from Bertelli et al. (2008):  $Z_{init}$ =0.017, $Y_{init}$=0.26, $\Delta Y / \Delta Z$ = 0, $Z_{\odot}/X_{\odot} $= 0.0165, except for $Z_{init}$=0.070 where $Y_{init}$=0.30 was used.}
\tablenotetext{b}{[Fe/H] calculated using values from Bertelli et al. (2008):  $Z_{init}$ =0.017, $Y_{init}$=0.26, $\Delta Y / \Delta Z$ = 2.25, $Z_{\odot}/X_{\odot} $= 0.0165, except for $Z_{init}$=0.070 where $Y_{init}$=0.30 was used.}
\tablenotetext{c}{[Fe/H] calculated using values from Mowlavi et al. (2012) for $\Delta Y / \Delta Z$ = 0, $Z_{\odot}/X_{\odot} $= 0.0174, and otherwise using the same parameters as in Bertelli et al. (2008):  $Z_{init}$ =0.017, $Y_{init}$=0.26, except for $Z_{init}$=0.070 where
$Y_{init}$=0.30 was used.}
\tablenotetext{d}{[Fe/H] calculated using values from Mowlavi et al. (2012) for $\Delta Y / \Delta Z$ = 1.2857, $Z_{\odot}/X_{\odot} $= 0.0174, and otherwise using the same parameters as in Bertelli et al. (2008):  $Z_{init}$ =0.017, $Y_{init}$=0.26, except for $Z_{init}$=0.070 where
$Y_{init}$=0.30 was used.}
\tablecomments{
We include the formula used for the conversion from Mowlavi et al. (2012), Equation A.4 for completeness: 
\begin{displaymath}
 \left[ \frac{Fe}{H} \right]_{init}=\log\left[ \frac{Z_{init}}{1 - Y_{init} - (1+ \Delta Y / \Delta Z) Z_{init} } \right] - \log \left[ \frac{Z_{\odot}}{X_{\odot}} \right]
\end{displaymath} 
}
\end{center}
\end{table}

\section{From Z to  [Fe/H]}
The metallicity parameter Z can be related to the [Fe/H] abundance, which is used observationally to characterize the metallicity of stars.  Table 3 displays a conversion to [Fe/H] from Z, where we have used the convenient formula from Mowlavi et al. (2012), which relates the initial Y and Z values to [Fe/H] given the galactic helium enrichment ratio $\Delta Y / \Delta Z$ and assumed values for the solar metallicity to hydrogen mass ratio $Z_{\odot} / X_{\odot}$.  

We adopted two sets of values for this calculation, based on the somewhat different parameters between  Bertelli et al. (2008) and Mowlavi et al. (2012).  We demonstrate that regardless of the helium enrichment ratio adopted, the values for [Fe/H] do not vary significantly.  For the second column in Table 1 we used $Z_{init}$ =0.017, $Y_{init}$=0.26, $\Delta Y / \Delta Z$ = 0, $Z_{\odot}/X_{\odot} $= 0.0165, except for $Z_{init}$=0.070 where $Y_{init}$=0.30 was used.   In the third column we used the He enrichment ratio from Bertelli et al. (2008), $\Delta Y / \Delta Z$ = 2.25.  Note the difference is just $+$0.03 dex at Z=0.017.  As a further test to see how sensitive the computed value for [Fe/H] was to the choice of parameters, we also used the parameters of  Mowlavi et al. (2012) for the solar metallicity to hydrogen ratio: $Z_{\odot}/X_{\odot} $= 0.0174, and as in column 2, using $\Delta Y / \Delta Z$ =0 .  Otherwise the same parameters as in the second column were adopted:  $Z_{init}$ =0.017, $Y_{init}$=0.26 and for $Z_{init}$=0.070 where $Y_{init}$=0.30.  For the fourth column we use the helium enrichment ratio $\Delta Y / \Delta Z$ = 1.2857, and we find the difference at Z=0.017 is only $+$0.01 dex.
 
We find that the metallicity [Fe/H] varies from $\sim$ $-$2 dex for the very metal poor star with Z=0.0001, to [Fe/H] $\sim$ $+$1 dex for a metal rich star with Z=0.070.  Note that for the solar metallicity case with Z=0.017, we calculate that [Fe/H] is 0.18, which is higher than one would expect it to be, since $[Fe/H]_{\odot}$ is by definition zero.  The reason for this is that this formula is for the \emph{initial} value, not the present day value for the solar metallicity.  Atomic diffusion is expected to cause the surface abundance to be reduced as the star evolves over time, as is discussed by Mowlavi et al. (2012).   We do not find much of a variation depending on the choice of the helium enrichment ratio or the choice of solar metallicity to hydrogen ratio.  The values differ at most by $\sim$ 0.1 dex, and only at the highest metallicity, Z=0.070, as seen in Table 1.  

The values in Table 1 correspond well to the range of metallicities observed in stars that host known exoplanets, which vary from about $-$0.6 dex to about $+$0.6 dex (e.g., Fischer \& Valenti 2005, Sousa et al. 2008, Ghezzi et al.  2010, and Buchhave et al. 2012).  We did not find a large difference in the duration of habitability between the cases at the extreme ends of metallicities, e.g., from Z=0.001 to Z=0.0001, or Z=0.040 to Z=0.070, as can be seen in Figs. 8-10.  

\begin{table}
\begin{center}
\caption{Location of Habitable Zone for  
duration greater than $10^9$ years\label{tbl-1}}
\begin{tabular}{cccccc}
\tableline\tableline
Z & 1.0 $M_{\odot}$ & 1.0 $M_{\odot}$ & 1.5 $M_{\odot}$   & 1.5  $M_{\odot}$& 2.0 $M_{\odot}$   \\
& Model A & Model B & Model A & Model B & Model A \\
 & (AU) & (AU) & (AU) & (AU)  & (AU) \\
\tableline
0.001 & 1.3-4.8 & 1.3-4.3  &  3.7-7.8 & 2.9-5.6  & ---  \\
0.017 & 0.8-3.3 & 0.8-3.3  & 2.2-5.9  &  2.1-5.6 & 4.3-9.2 \\
0.040 & 0.6-2.6 & 0.7-2.7  & 1.7-4.6  &  1.7-4.6 & 3.1-8.0\\
\tableline
\end{tabular}
\tablecomments{
Model A indicates that the simple calculation of the planet temperature of Eq. (1) was used to estimate the inner and outer boundaries of the habitable zone.  Model B indicates that the more refined calculation of the inner and outer boundaries of the habitable zone, Eqs. (2) and (3), were used.
}
\end{center}
\end{table}

\begin{figure}
\plotone{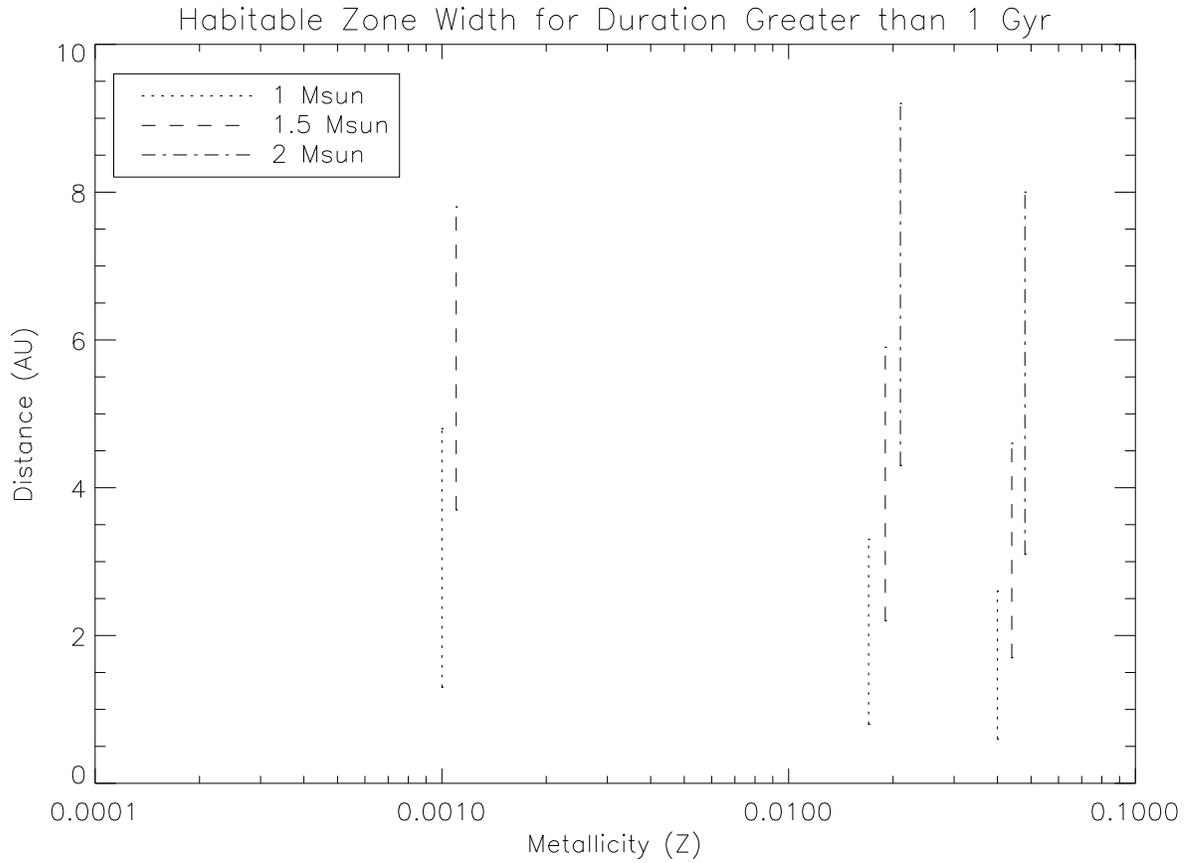}
\caption{Habitable Zone location versus metallicity for durations greater than 1 Gyr for stars with initial masses 1.0, 1.5, and 2.0 $M_{\odot}$.  Note that for low metallicity stars (Z$ \leqslant$ 0.001) with initial mass 2.0 $M_{\odot}$ the duration is shorter than 1 Gyr and therefore is not included in this plot. }{\label{fig18}}
\end{figure}

\section{Discussion}
The fraction of stars in the solar neighborhood by MK spectral classes for those with magnitudes brighter than V=8.5 from the HD catalog is 22\% for A stars, 19\% for F stars, and 14\% for G stars out of a total of about $10^5$ stars in the catalog (Cox et al. 2000a).  The choice of initial masses for this paper correspond to MK spectral classes G2V for 1.0 $M_{\odot}$, F2.5V for 1.5 $M_{\odot}$, and A5V for 2.0 $M_{\odot}$ (Cox et al. 2000b), so the results of this paper are relevant to over half the stars in the solar neighborhood.

A  summary of the results from our calculations is displayed in Table 2 and Fig. 18 for Z=0.001, 0.017, and 0.040, respectively, for the 1.0, 1.5, and 2.0 $M_{\odot}$ stars, assuming it takes $10^9$ years for life to begin, to develop, and to become observable.  The results for a 2.0 $M_{\odot}$ initial mass star with Z=0.001 are not included in the figure because the duration of habitability is only about $6 \times 10^8$ years.   

The relatively short duration of $\sim$ $10^8$ years is also of interest, and it occurs during the core He burning phase, as seen in Figs. 8-10 and 16-17.  If panspermia operates between planets, such as between Mars and the Earth, as suggested by Davies (2003), then it is possible for life to be seeded and be observable for outer planets at distances of 5 to 15 AU during that time period, assuming biogenesis occurs on the shortest durations of the order of tens of millions of years, as discussed by Lineweaver and Davis (2002), and as seen in their Figs. (2) and (3).

\begin{table}
\begin{center}
\caption{Location of Habitable Zone for  
duration greater than $3 \times 10^9$ years\label{tbl-2}}
\begin{tabular}{cccccc}
\tableline\tableline
Z & 1.0 $M_{\odot}$  & 1.0 $M_{\odot}$ & 1.5 $M_{\odot}$ &1.5 $M_{\odot}$  & 2.0 $M_{\odot}$   \\
  &  Model A &  Model B &  Model A & Model B& Model A \\
  & (AU) & (AU) &  (AU) & (AU) & (AU) \\
\tableline
0.001 &  1.6-3.6 &  1.5-3.2  &   ---        & ---         & ---  \\
0.017 &  0.9-2.8 &  0.9-2.8 &  2.6-5.0 & 2.6-4.5  & --- \\
0.040 &  0.7-2.3 &  0.7-2.4 &  1.9-4.3 &  1.9-4.0 & --- \\
\tableline
\end{tabular}
\tablecomments{
Model A indicates that the simple calculation of the planet temperature of Eq. (1) was used to estimate the inner and outer boundaries of the habitable zone  were.  Model B indicates that the more refined calculation of the inner and outer boundaries of the habitable zone, Eqs. (2) and (3), were used.
}
\end{center}
\end{table}

\begin{figure}
\plotone{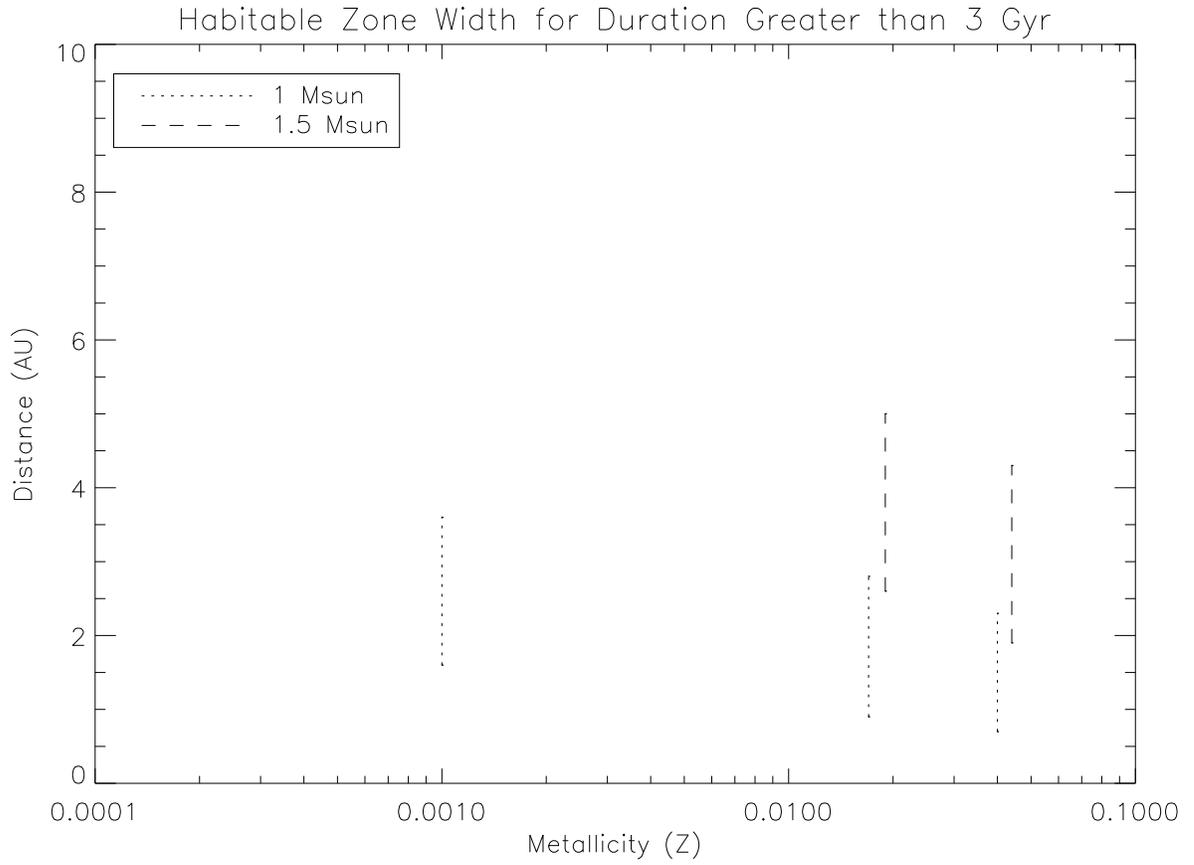}
\caption{Habitable Zone location versus metallicity for durations greater than 3 Gyr for stars with initial masses 1.0 and 1.5  $M_{\odot}$.  Note that for  stars with initial mass 2.0 $M_{\odot}$ the duration is shorter than 3 Gyr for all metallicities and therefore is not included in this plot. }{\label{fig19}}
\end{figure}

However, the duration necessary for life to develop and to become observable could be much longer than either time scale mentioned so far.  In their studies of habitable stars for the SETI project, Turnbull and Tarter (2003) adopted $3 \times 10^9$ years.  Their reason for choosing the longer duration  was that multicellular life did not appear until after $3 \times 10^9$ years based on the fossil record (Rasmussen et al. 2002) and on biomolecular clocks (Ayala, Rzhetsky, \& Ayala 1998, and Wray, Levinton, \& Shapiro 1996).  In Fig. 19 and Table 3 we present our results for this case.  We find that the width of the habitable zone decreases substantially for 1.0 and 1.5 $M_{\odot}$ stars for all metallicities.  For 1.5 $M_{\odot}$ stars, only the higher metallicity cases, Z=0.017 and 0.040 have durations greater than $3 \times 10^9$ years, and for 2.0 $M_{\odot}$ stars, the duration is always shorter than $3 \times 10^9$ years, and hence is not shown in the figure.

In this paper, we identify three time scales for further study, available for A, F, and G stars from 1.0 to 2.0 solar masses in the solar neighborhood. These are:  short, 100 Myr, medium, 1 Gyr, and long, 3 Gyr, which are potentially observable.  These time scales allow observational tests of concepts of hard step scenarios of evolution (Lineweaver and Davis, 2003, Carter 2008), particularly if the first one to three steps are hard, since they cover time periods of up to about 2.3 Gy from the formation of the Earth (Carter 2008).  By 3 Gyr there should be changes in atmospheric composition due to life, as seen in  Fig. 1 of Carter (2008). 

 Our results imply that we can use a combination of age and metallicity  to determine the timescale for the observability of life--whether life evolves and affects a planet's atmosphere on relatively short timescales, of the order of $10^9$ years, or several times longer.  In all cases the longest lifetimes are for the highest metallicity stars.  Also, high metallicity stars have habitable zones located closer to the stars themselves due to their lower luminosity.  Importantly, for each star of a given initial mass, there is an optimum region for habitability.  

\section{Summary and Conclusions}
During the course of the evolution of a star, the location and width of the habitable zone changes as its luminosity and radius evolves.  The duration of habitability for a planet located at a given distance from a star is greatly affected by the characteristics of the host star, and it affects the conditions for the development of life.  The longer the duration of habitability, the more likely it is that life has evolved.   Knowledge of the distance from the star and the conditions of habitability including its duration  need to be quantified  in order to prepare the observational technique(s)  that can be optimally used to detect life.   

We calculated the evolution of the habitable zone around stars of 1.0, 1.5, and 2.0 solar masses for metallicities ranging from a Z=0.0001 to Z=0.070.  These calculations include the evolution of the habitable zone  from the PMS until the AGB is reached.   Our new study confirms our previous work (Lopez, Schneider, \& Danchi 2005) and greatly extends it.  We find that metallicity strongly affects the duration of the habitable zone for a planet as well as distance from the host star where the duration is maximized.  For a 1 solar mass  star with near Solar metallicity (Z=0.017) we find that the duration of the habitable zone is greater than $10^{10}$ years from 1.2 to 2.0 AU from the star, whereas duration is greater than $2 \times 10^{10}$ years for high metallicity stars (Z=0.070) at distances of 0.7 to 1.8 AU, and much lower, about  $4 \times 10^{9}$ years, for low metallicity stars (Z=0.0001).   Corresponding results have been obtained for stars of 1.5 and 2.0 solar masses.  

Our results have observational implications:  given a broad range in metallicities and stellar masses for stars in the solar neighborhood, it is possible to determine the time scale for life to evolve and to be observable, whether it is less than or equal to 1 billion years, or much longer, such as 3 billion years.   In a few years, we will have a sample of nearby stars with known planets in the habitable zone.  If a sufficiently large sample is found, and if we can determine stellar ages with sufficient precision, hopefully we will be able to detect life and determine the characteristic time scale for its development.

\acknowledgments
We are grateful to J. Scalo for having suggested further work into the habitable zone of stars after the He flash and to J.-L. Menut for useful and enthusiastic discussions about this work.  We thank L. Danchi for a careful reading of the manuscript and suggestions for improvement.


\clearpage

\end{document}